%% file: paper.tex
\documentclass[conference]{IEEEtran}
\usepackage{cite}
\usepackage{amsmath,amssymb,amsfonts, amsthm}
\usepackage{algorithm}
\usepackage{algorithmicx}
\usepackage[noend]{algpseudocode}
\usepackage{graphicx}
\usepackage{textcomp}
\usepackage[dvipsnames]{xcolor}

\usepackage{etex}
\usepackage{hyperref}
\usepackage[utf8]{inputenc}
\usepackage[T1]{fontenc}
\usepackage{wrapfig}
\usepackage{times}
\usepackage{paralist}
\usepackage{subfig}
\usepackage{booktabs}
\usepackage{paralist}
\usepackage{amssymb, amsfonts}
\usepackage{amsmath}
\usepackage{amsopn}
\usepackage{url}
\usepackage{multirow}
\usepackage{relsize}
\usepackage{xfrac}
\usepackage{pdflscape}
\usepackage{tabularx}
\usepackage{xspace}
\usepackage{balance}
\usepackage{contour}
\usepackage[normalem]{ulem}
\usepackage{nicefrac}
\usepackage{mdframed}
\usepackage{marginnote}
\usepackage{comment}
\usepackage{diagbox}
\usepackage{cleveref}

\contourlength{0.8pt}
\newcommand{\myuline}[1]{%
	\uline{\phantom{#1}}%
	\llap{\contour{white}{#1}}%
}

\newcommand{\mypar}[1]{\smallskip\noindent\textbf{#1.}}
\newcommand{\mypartwo}[1]{\noindent\textit{#1.}}

\def\Snospace~{\S{}}

\newcommand{\eventlog}{\emph{L}\xspace}
\newcommand{\abstractedlog}{\emph{L'}\xspace}

\newcommand{\eventclassesinlog}{C_L\xspace}
\newcommand{\finalgrouping}{\emph{G}\xspace}
\newcommand{\optimalgrouping}{\emph{$\hat{G}$}\xspace}
\newcommand{\allcandidates}{\mathcal{G}\xspace}
\newcommand{\constraints}{\mathcal{R}\xspace}
\newcommand{\userconstraints}{\emph{R}\xspace}
\newcommand{\dfg}{DFG\xspace}
\newcommand{\approach}{\textsc{gecco}\xspace}

\newcommand{\abstractf}{\texttt{abstract}\xspace}
\newcommand{\dist}{\texttt{dist}\xspace}
\newcommand{\inst}{\texttt{inst}\xspace}
\newcommand{\holds}{\texttt{holds}\xspace}
\newcommand{\interrupts}{\texttt{interrupts}\xspace}
\newcommand{\missing}{\texttt{missing}\xspace}
\newcommand{\gclrkone}{g_{\emph{clrk1}}\xspace}
\newcommand{\gclrktwo}{g_{\emph{clrk2}}\xspace}
\newcommand{\clrkone}{\emph{clrk1}\xspace}
\newcommand{\clrktwo}{\emph{clrk2}\xspace}
\newcommand{\gmgr}{g_{\emph{mgr}}\xspace}
\newcommand{\configexh}{\textsc{e}xh\xspace}
\newcommand{\configdfg}{\textsc{dfg}$_\infty$\xspace}
\newcommand{\configdfgbeam}{\textsc{dfg}$_{k}$\xspace}
\newcommand{\solved}{Solved\xspace}
\newcommand{\sizered}{S. red.\xspace}
\newcommand{\complred}{C. red.\xspace}
\newcommand{\silhouette}{Sil.\xspace}
\newcommand{\runtime}{T(m)\xspace}
\newcommand{\blgreedy}{\textsc{bl$_{G}$}\xspace}
\newcommand{\blpart}{\textsc{bl}$_{\mathit{P}}$\xspace}
\newcommand{\blquery}{\textsc{bl}$_{\mathit{Q}}$\xspace}

\newcommand{\true}{\ensuremath{\mathit{true}}\xspace}
\newcommand{\false}{\ensuremath{\mathit{false}}\xspace}

\theoremstyle{definition}
\newtheorem{problem}{Problem}

\newcommand{\mytitle}{
	\approach:
Constraint-driven Abstraction of \\ Low-level Event Logs
}

\def\BibTeX{{\rm B\kern-.05em{\sc i\kern-.025em b}\kern-.08em
    T\kern-.1667em\lower.7ex\hbox{E}\kern-.125emX}}

\hypersetup{
	bookmarks=true,        
	unicode=true,          
	pdffitwindow=false,    
	pdfstartview={FitH},   
	pdftoolbar=true,       
	pdfmenubar=true,       
	colorlinks=true,      
	linkcolor={Brown},       
	urlcolor={Brown},        
	citecolor={Brown},       
	filecolor=black,       
	pdftitle={\mytitle},           
	pdfauthor={Rebmann et al.}, 
	pdfcreator={Rebmann et al.} 
}
\begin{document}

\title{\mytitle}

\author{\IEEEauthorblockN{Adrian Rebmann}
\IEEEauthorblockA{
\textit{University of Mannheim, Germany}\\
rebmann@informatik.uni-mannheim.de}
\and
\IEEEauthorblockN{Matthias Weidlich}
\IEEEauthorblockA{
\textit{Humboldt-Universit\"at zu Berlin, Germany}\\
matthias.weidlich@hu-berlin.de}
\and
\IEEEauthorblockN{Han van der Aa}
\IEEEauthorblockA{
\textit{University of Mannheim, Germany}\\
han@informatik.uni-mannheim.de}}

\maketitle

\begin{abstract}
	\input{sections/abstract.tex}
\end{abstract}


\section{Introduction}
\label{sec:introduction}
\input{sections/introduction.tex}

\section{Motivation}
\label{sec:motivation}
\input{sections/motivation.tex}

\section{Problem Statement}
\label{sec:problem}
\input{sections/problem.tex}

\section{\approach: Scope}
\label{sec:scope}
\input{sections/scope.tex}

\section{The \approach Approach}
\label{sec:approach}
\input{sections/approach.tex}

\section{Experimental Evaluation}
\label{sec:eval}
\input{sections/evaluation.tex}

\section{Related Work}
\label{sec:related}
\input{sections/related.tex}

\section{Conclusion}
\label{sec:conclusion}
\input{sections/conclusion.tex}

\bibliographystyle{IEEEtran}
\bibliography{IEEEabrv,refs}

\balance

\end{document}

%% file: sections/abstract.tex
%
Process mining enables the analysis of complex systems using event data 
recorded during the execution of processes. Specifically, models of 
these processes can be discovered from event logs, i.e., sequences of events. 
However, the recorded events are often too fine-granular 
and result in unstructured models that are not meaningful for 
analysis. Log abstraction therefore aims to group together events to obtain a 
higher-level representation of the event sequences. While such a transformation 
shall be driven by the analysis goal, existing techniques force 
users to define 
\emph{how} the abstraction is done, rather than \emph{what} the result 
shall be. 

In this paper, we propose \approach, an approach for log abstraction 
that enables users to impose requirements on the resulting log in terms of 
constraints. \approach then groups events so that the constraints are 
satisfied and the distance to the original log is minimized. 
Since exhaustive log abstraction suffers from an exponential runtime 
complexity, 
\approach also offers a 
heuristic approach guided by behavioral dependencies found in the 
log. We show that the abstraction quality of 
\approach is superior to baseline solutions and demonstrate the relevance of considering constraints during log abstraction in real-life settings.

%% file: sections/introduction.tex

Process mining~\cite{van2016process} comprises methods to analyze 
complex systems based on event data that is recorded during the execution of 
processes. Specifically, by discovering models of 
these processes from event logs~\cite{augusto2019discovery}, i.e., sequences of 
events, process mining yields insights into how a process is truly 
executed. 
Yet, the recorded events are often too fine-granular for meaningful 
analysis, and the resulting variability in the recorded event sequences leads 
to complex models. This trend is amplified when events originate from 
sources, such as real-time location 
systems~\cite{SenderovichRGMM16} and user interface 
logs~\cite{leno2020identifying}.

To tackle this issue, \emph{log abstraction} is a technique to 
lift the event sequences of a log to a more abstract representation, by 
grouping low-level events into high-level activities. Existing techniques for 
log abstraction (cf., \cite{VanZelst2020,Diba2020}) differ in the adopted 
algorithms and employ, e.g., temporal clustering of events~\cite{DeLeoni2020} 
or the detection of predefined 
patterns~\cite{mannhardt18patterns}. Yet, their focus is on \emph{how} the 
abstraction is conducted, rather than \emph{what} properties the abstracted log 
shall satisfy. Without control on the result of log abstraction, 
however, it is hard to ensure that an abstraction is appropriate for a specific 
analysis goal.

To achieve effective abstraction, a respective technique 
must thus enable users to incorporate dedicated constraints on the resulting log. 
Here, the main challenge is that such constraints may be defined at different 
levels of granularity, i.e., they may relate to properties of individual 
events, types of events, or groups of event types. Finding an optimal 
abstraction, i.e., a log that satisfies all constraints while being as 
close as possible to the original log, is a hard problem, due to the sheer 
number of possible abstractions and the interplay of constraints at 
different granularity levels. Hence, log abstraction is challenging also from a 
computational point of view.

In this paper, we propose \approach, an approach for log abstraction that 
enables a user to impose requirements on the resulting log in terms of 
constraints. As such, it supports a declarative characterization of the 
properties the abstracted log shall adhere to, in order to be meaningful for a 
specific analysis purpose. We summarize our contributions as follows:
\begin{compactitem}
\item We define the problem of optimal log abstraction. It requires 
minimizing the distance to the original log while satisfying a set of 
constraints on the abstracted log.
\item We define the scope of \approach as an instantiation of the 
log-abstraction problem, covering a broad set of common constraint types 
and a distance measure.
\item As part of \approach, we present an algorithm for 
exhaustive log abstraction. Striving for more efficient processing, we also 
provide a heuristic algorithm that is guided by behavioral dependencies found 
in the log.
\end{compactitem}
Our evaluation demonstrates that 
the abstracted logs obtained with \approach provide better abstraction and are 
more cohesive than those 
obtained with baseline techniques. As such, process discovery algorithms also 
yield more structured models.

In the remainder, we first motivate the need for user-defined constraints in 
log abstraction (\autoref{sec:motivation}). We then formally define the problem 
of optimal log abstraction (\autoref{sec:problem}). Next, \approach is 
presented as an instantiation of the log-abstraction 
problem (\autoref{sec:scope}), along with exhaustive and heuristic algorithms 
to 
address it (\autoref{sec:approach}).
Finally, we report on evaluation results (\autoref{sec:eval}), review related 
work (\autoref{sec:related}), and conclude (\autoref{sec:conclusion}).

%% file: sections/motivation.tex

\begin{figure}[!htb]
	\centering
	\includegraphics[width=\linewidth]{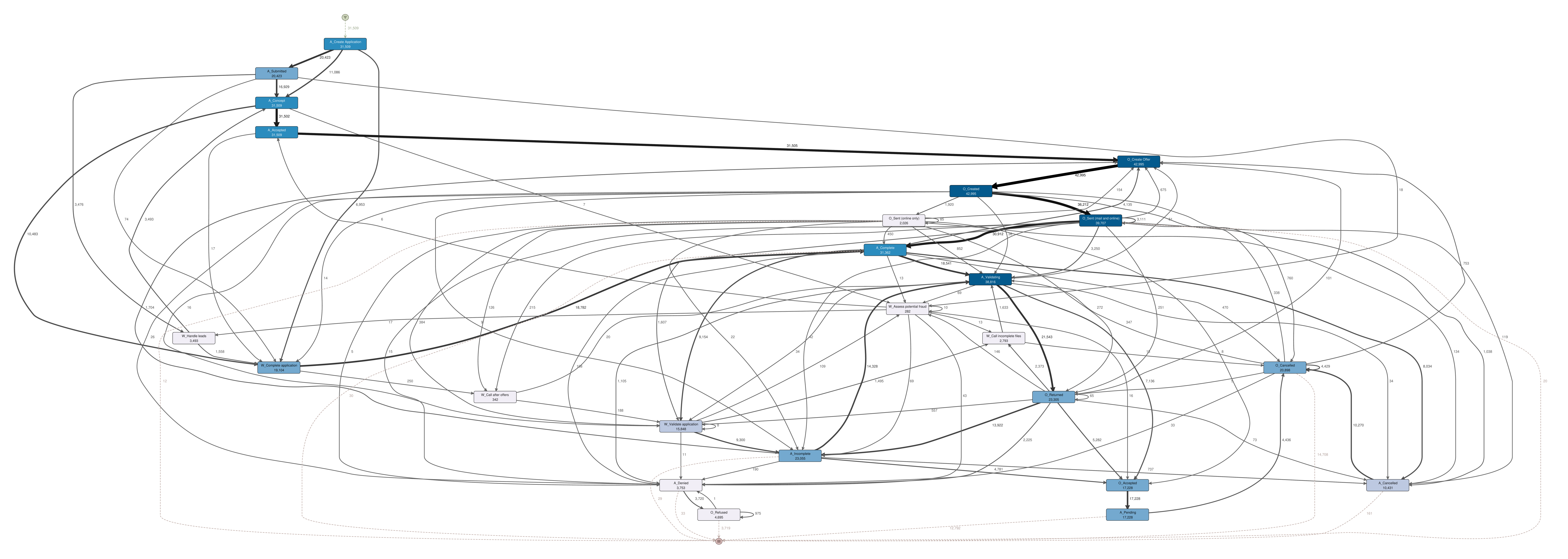}
	\vspace{-.8em}
	\caption{A so-called \emph{spaghetti} process model.}
	\label{fig:spaghetti}
\end{figure}

Log abstraction is motivated by the presence of fine-granular events in process mining settings. 
Such fine-granular events 
typically induce a high degree of behavioral variability, so that the application of process discovery algorithms yields so-called 
\emph{spaghetti process 
	models}, which are incomprehensible due to their complexity, 
as e.g., depicted in \autoref{fig:spaghetti}.
Log abstraction overcomes this issue by grouping events, thereby 
reducing the variability of the behavior to be depicted. 

For illustration purposes, consider 
the simple event log in \autoref{tab:runningexample}, which consists of four 
event sequences 
corresponding to a request-handling process. Here, 
\underline{\textcolor{blue}{blue underlined events}} denote process 
steps performed by a clerk, whereas the others are performed by a manager. 

The event log shows that each case starts with the receipt of a request (event \emph{rcp}) by a clerk. 
The clerk checks the request either casually (\emph{ckc}) or thoroughly 
(\emph{ckt}) depending on the information provided. 
Then, the request is forwarded to a manager, who either accepts (\emph{acc}) or 
rejects (\emph{rej}) it. Afterwards, the clerk may or may not assign priority 
to a request (\emph{prio}), before they inform the customer (\emph{inf}) and 
archive the request (\emph{arv}). The latter two activities can be 
performed in either order, as shown, e.g., in $\sigma_1$ and $\sigma_2$. As 
shown in trace $\sigma_{4}$, a rejected request may also be returned to the 
applicant, who will resubmit it, restarting the procedure.

\begin{table}[!htb]
	\small
	\centering
	\caption{Exemplary traces of an event log.}
	\label{tab:runningexample}
	\begin{tabular}{cl}
		\toprule
		\textbf{ID} & \textbf{Trace} \\
		\midrule
		$\sigma_1$ & $\langle$\textcolor{blue}{\myuline{rcp}}, \textcolor{blue}{\myuline{ckc}}, acc, \textcolor{blue}{\myuline{prio}}, \textcolor{blue}{\myuline{inf}}, \textcolor{blue}{\myuline{arv}}$\rangle $\\
		$\sigma_2$ &  $\langle$\textcolor{blue}{\myuline{rcp}}, \textcolor{blue}{\myuline{ckt}},  rej, \textcolor{blue}{\myuline{prio}}, \textcolor{blue}{\myuline{arv}}, \textcolor{blue}{\myuline{inf}}$\rangle $\\
		$\sigma_3$ &  $\langle$\textcolor{blue}{\myuline{rcp}}, \textcolor{blue}{\myuline{ckc}}, acc, \textcolor{blue}{\myuline{inf}}, \textcolor{blue}{\myuline{arv}}$\rangle$\\
		$\sigma_4$ & $\langle$\textcolor{blue}{\myuline{rcp}},  \textcolor{blue}{\myuline{ckc}}, rej, \textcolor{blue}{\myuline{rcp}}, \textcolor{blue}{\myuline{ckt}}, acc, \textcolor{blue}{\myuline{prio}}, \textcolor{blue}{\myuline{arv}},\textcolor{blue}{\myuline{inf}}$\rangle$ \\
		\bottomrule
	\end{tabular}
\end{table}

\begin{figure}[!htb]
	\centering
	\includegraphics[width=0.8\linewidth]{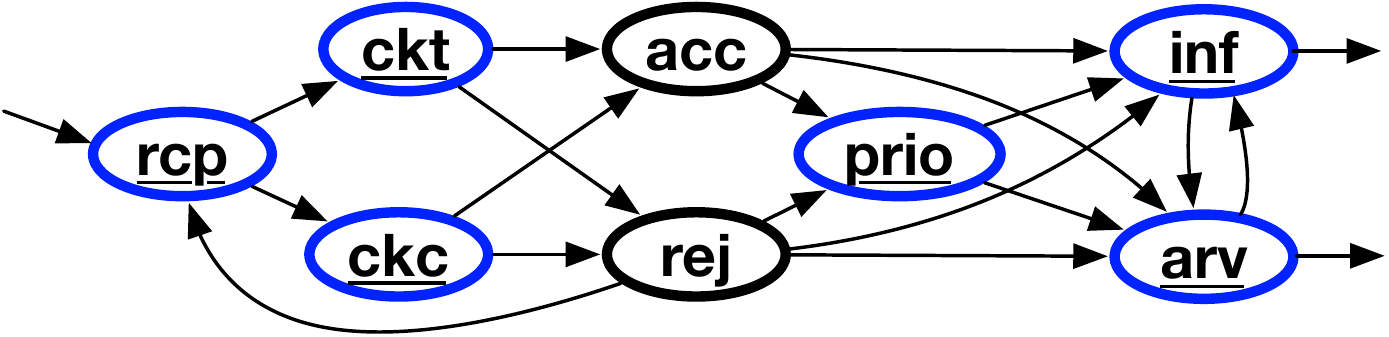}
	\caption{Directly-follows graph (DFG) of the running example.}
	\label{fig:initialdfg}
	\vspace{-.5em}
\end{figure}

Although this process consists of only eight distinct steps, its 
behavior is already fairly complex. This is evidenced by the 
\emph{directly-follows graph} (DFG) shown in \autoref{fig:initialdfg}, which 
depicts the steps that can directly succeed each other in the process.
The graph's complexity already obscures some of the key 
behavioral aspects of the process. Log 
abstraction may alleviate this problem. However, existing techniques focus on 
how the abstraction shall be done. For instance, they may exploit that 
the steps \emph{ckt}, 
\emph{ckc}, \emph{acc}, and \emph{rej} are closely correlated from a behavioral 
perspective and abstract them to a single activity. Yet, this is not meaningful 
for many analysis tasks, as it would obscure the fact 
that the activity encompasses some steps performed by a clerk 
(\emph{ckt} and \emph{ckc}), whereas others are performed by a manager 
(\emph{acc} and \emph{rej}).

By incorporating user-defined constraints on what properties the abstracted log 
shall satisfy, such a result can be avoided.
For instance, if a user wants to primarily understand the interactions between employees, while abstracting from details on the individual steps performed by them, a constraint may enforce that each 
activity comprises only events performed by the same employee role.
If applied in a naive manner, this constraint would result in two groups of 
events classes, i.e., $g_{\emph{clrk}} = \{ \emph{rcp}, \emph{ckc}, \emph{ckt}, 
\emph{prio}, \emph{inf}, \emph{arv}\}$ and $\gmgr = \{\emph{acc}, 
\emph{rej}\}$. Yet, using these groups directly for log abstraction is not 
meaningful either. The group $g_{\emph{clrk}}$ includes 
steps that occur at the start of the process, as well as steps that only happen 
at the end. Moreover, abstracting the steps in $\gmgr$ to a single activity 
would obfuscate that $\{\emph{acc}, 
\emph{rej}\}$ exclude each other and that only after step $\emph{rej}$, the 
process is potentially restarted.  

\begin{figure}[!htb]
	\centering
	\includegraphics[width=0.5\linewidth]{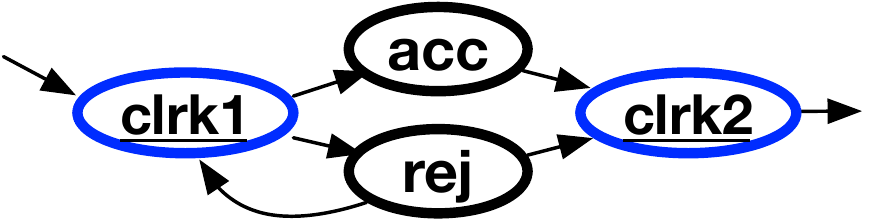}
	\caption{DFG of the log abstracted with \approach.}
	\label{fig:abstracteddfg}
	\vspace{0em}
\end{figure}

Against this background, our approach to log abstraction, \approach, aims at 
constructing activities for groups of events that satisfy user-specified 
constraints, while also preserving the behavior represented in event sequences 
as much as possible. 
For the example, this would result in an abstraction that consists of four 
groups: $\gclrkone = \{\emph{rcp}, \emph{ckc}, \emph{ckt}\}$, containing the 
initial steps performed by the clerk, $\{\emph{acc}\}$ and $\{\emph{rej}\}$, as 
singleton groups of steps that are mutually exclusive and both performed 
by the manager, and $\gclrktwo = 
\{\emph{prio}, \emph{inf}, \emph{arv}\}$, the final steps of the process 
performed by the clerk. The directly-follows graphs obtained with this log 
abstraction is shown in \autoref{fig:abstracteddfg}. It highlights that a clerk 
starts working on each case, before handing it over to the manager. Accepted 
requests are completed by the clerk, whereas rejected requests may be completed 
or returned to the start of the process.

Obtaining the groups for such abstraction requires solving a 
computationally complex problem, though, as defined next.

%% file: sections/problem.tex

We first define an event model for our work (\autoref{sec:preliminaries}), 
before formalizing the addressed log-abstraction problem 
(\autoref{sec:problemformulation}).

\subsection{Event model}
\label{sec:preliminaries}

We consider events recorded during the execution of a process and write 
$\mathcal{E}$ for the universe of all events.
An event $e \in \mathcal{E}$ is of a certain \emph{event class}, i.e., its 
type, which we denote as $e.C \in \mathcal{C}$, with $\mathcal{C}$ as the 
universe of event classes. For instance, the log in 
\autoref{tab:runningexample} consists of eight event classes, each 
corresponding to a specific process step.
Furthermore, each event carries information about its context, which may 
include aspects such as a timestamp, the executing role, or relevant data 
values. 
We capture this context by a set of data attributes $\mathcal{D} = \{ D_1, 
\ldots, D_p \}$, with $dom(D_i)$ as the domain of attribute $D_i$, $1 \leq i 
\leq p$. We write $e.D$ for the value of attribute $D$ of an event $e$. 
For instance, in \autoref{sec:motivation}, each event class is assigned a 
particular role, i.e., a 
\emph{clerk} or a \emph{manager}. As such, an event $e$ may capture that $e.C = 
\emph{rcp}$ and $e.\mathit{role} = \emph{clerk}$.

A single execution of a process, called a \emph{trace}, is modeled as a 
sequence of events $\sigma \in \mathcal{E}^*$, such that no event can occur 
in more than one trace. An event log is a set of traces, $L \subseteq 
2^{\mathcal{E}^*}$, with $\mathcal{L}$ as the universe of all events logs, and $C_L \subseteq \mathcal{C}$ as the set of event classes of the events in $L$.

An event log can be represented as a \emph{directly-follows 
	graph} (\dfg) that indicates if two event classes 
ever immediately succeed each other in the log. 
Given a log $\eventlog$, its \dfg is a directed graph $(V, E)$, with the set of vertices $V$ 
corresponding to the event classes of $C_L$ and the set of edges $E \subseteq V \times V$ 
representing a directly-follows relation $>_{\eventlog}$, defined as: 
$a >_{\eventlog} b$, if there is a trace $\sigma =\langle e_1,
\ldots,e_n\rangle$ and $i \in \{1,\ldots,n-1\}$, such that $\sigma \in 
\eventlog$ and $e_i.C = a$ and $e_{i+1}.C = b$. 

\subsection{The log-abstraction problem}
\label{sec:problemformulation}

Log abstraction aims to construct groups of  
similar events for an event log. 
Formally, this is captured by a grouping, i.e., a set of groups, 
$\finalgrouping \subset 2^\mathcal{C}$, over the event classes 
$\eventclassesinlog$, such that each class $c \in C_L$ is part of exactly one 
group $g \in \finalgrouping$.
Given a grouping, a function $\abstractf: \mathcal{L} \times 2^\mathcal{C} 
\rightarrow \mathcal{L}$ is applied to obtain an abstracted log $L'$.  For 
instance, using the log and grouping from \autoref{sec:motivation}, 
$\sigma_1$ is abstracted to $\sigma'_1 = \langle \clrkone, \emph{acc}, \clrktwo 
\rangle$. 

We target scenarios in which a user formulates requirements on what properties 
the abstracted event log and, hence, the grouping $\finalgrouping$ shall 
satisfy, e.g., to group only events performed by a single role (see 
\autoref{sec:motivation}). 
Then, we aim to identify a grouping $\optimalgrouping$ 
that meets these requirements, while preserving the behavior of the traces as 
much as possible. To this end, we define $\dist: \mathcal{L} \times 
2^\mathcal{C} \rightarrow \mathbb{R}$ as a \emph{distance function} that 
quantifies the distance of a grouping to an event log. 
Also, using 
$\mathcal{R}$ to denote the universe of possible constraints, we define a 
predicate $\holds: 2^\mathcal{C} \times \mathcal{R} \times \mathcal{L} 
\rightarrow \{\true, \false\}$ to denote whether a grouping satisfies a set of 
constraints for a given log. Based thereon, we define the optimal 
log-abstraction 
problem:

\begin{problem}[Optimal log abstraction]
	\label{problem}
	Given an event log $L$ with event classes $C_L$, a distance function 
	$\dist$, and a set of constraints $\userconstraints$, the log-abstraction 
	problem is to find an optimal grouping $\optimalgrouping = \{g_1, \ldots, 
	g_k \}$, such that: 
	\begin{compactitem}
		\item $\optimalgrouping$ is an exact cover of $\eventclassesinlog$, 
		i.e., $\bigcap_{i}^{k}g_i=\emptyset \land 
		\bigcup_{i}^{k}g_i=\eventclassesinlog$;
		
		\item $\optimalgrouping$ adheres to the desired constraints $\userconstraints$, i.e., 
		$\holds(\optimalgrouping, \userconstraints, \eventlog) = true$;
				
		\item the distance $\dist(\optimalgrouping,\eventlog)$ is minimal.
	\end{compactitem}
	
\end{problem}

%% file: sections/scope.tex

To address \autoref{problem}, we propose \approach for the \underline{G}rouping 
of \underline{E}vent \underline{C}lasses using \underline{C}onstraints and 
\underline{O}ptimization. This section shows how \approach instantiates 
\autoref{problem} by specifying constraint types 
(\autoref{sec:scope:constraints}) and a distance function 
(\autoref{sec:scope:distance}). 

\subsection{Covered constraint types}
\label{sec:scope:constraints} 

\approach is able to handle a broad range of constraints on a grouping $\finalgrouping$. 
As shown through the examples in \autoref{tab:constraints}, we consider 
\emph{grouping} constraints, \emph{class-based} constraints, and 
\emph{instance-based} constraints. 
The table also indicates a monotonicity property of the constraints, which is 
important when aiming to find an optimal grouping in an efficient manner.


\begin{table*}[!htb]
	\centering
	\caption{Exemplary constraints covered by \approach.}
	\label{tab:constraints}	
	\begin{tabular}{lll}
		\toprule
		\textbf{Category}& \textbf{Examples} & \textbf{Monotonicity}\\		
		\midrule
		\multirow{2}{2cm}{Grouping\\constraints}&There should be at most 10 groups in the final grouping.& \multicolumn{1}{c}{n/a}\\
		&There should be at least 5 groups in the final grouping.& \multicolumn{1}{c}{n/a}\\
		\midrule 
		\multirow{4}{2cm}{Class-based\\constraints} & There should be at least 5 event classes per group.& monotonic\\
				&At most 10 event classes should be grouped together. &anti-monotonic\\
		&The event classes \emph{rcp} and \emph{acc} cannot be members of the same group.&anti-monotonic\\
		&The event classes \emph{inf} and \emph{arv} must be members of the same group.&non-monotonic\\
		\midrule 
		\multirow{6}{2cm}{Instance-based\\constraints}&
		At least 2 distinct document codes must be associated with a group instance.&monotonic\\
		&The cost of a group instance must be at most 500\$.&anti-monotonic\\
		&The duration of a group instance must be at most 1 hour on average.&non-monotonic\\
		&The time between consecutive events in a group instance must at most be 10 minutes.&anti-monotonic\\
		&Each group instance may contain at most 1 event per event class. & anti-monotonic \\
		&At least 95\% of the group instances must have a cost below 500\$. &anti-monotonic\\
		\bottomrule 
	\end{tabular}
	\vspace{-1em}
\end{table*}

\mypar{Grouping constraints} This constraint category can be used to bound the size of  a grouping $\finalgrouping$, i.e., 
the number of high-level activities that will appear in the abstracted log. 
An upper bound restricts the size and complexity of the obtained log, whereas a lower bound 
can limit the applied degree of abstraction. 

\mypartwo{Satisfaction}
We use $R_G \subseteq R$ to refer to the subset of grouping constraints. 
Whether a constraint $r \in R_G$ holds can be directly checked against the 
grouping size, $|\finalgrouping|$. As such, for the \holds 
predicate, we require $\forall r \in 
\userconstraints_\finalgrouping: r(\finalgrouping)=\true$.


\mypar{Class-based constraints} The second category of constraints can be used to influence
the characteristics of an individual group $g \in \finalgrouping$ in terms of the event classes that it can contain.
\approach supports any class-based constraint for which satisfaction can be checked by considering $g$ in isolation, i.e., without having to compare $g$ to other groups in \finalgrouping. As shown in \autoref{tab:constraints}, this, for instance, includes constraints 
that each group  shall comprise at least (or at most) a certain number of 
event classes, as well as \emph{cannot-link} and \emph{must-link} constraints, which may be used to specify that two event classes must or must not be grouped together.

\mypartwo{Satisfaction}
We use $R_C \subseteq R$ for  class-based constraints. The 
satisfaction of a constraint $r \in R_C$ 
is directly checked by evaluating the contents of each group $g \in 
\finalgrouping$. Hence, the 
\holds predicate requires that $\forall g\in \finalgrouping, \forall r \in 
\userconstraints_C : r(g) = \true$. 

\mypartwo{Monotonicity}
Class-based constraints that specify a minimum requirement on groups, e.g., a 
minimal group size, are monotonic: If the constraint holds for a group $g$, it 
also holds for any larger group $g'$, with $g \subset g'$. In other words, 
adding event classes to a group can never result in a (new) constraint 
violation.
By contrast, constraints that express requirements that may not be exceeded, 
e.g., a maximal group size or \emph{cannot-link} constraint, are 
anti-monotonic: If they hold for a group $g$, they also hold for any subset of 
that group $g' \subset g$. However, if a group $g$ violates a constraint, a 
larger group $g'$, with $g \subset g'$, also violates it.

\mypar{Instance-based constraints} The third category comprises constraints 
that shall hold for each \emph{instance} of 
a group $g \in \finalgrouping$, i.e., a sequence of (not necessarily 
consecutive) events that occur in the same trace and of which the event classes 
are part of $g$.
In line with the event context defined in 
\autoref{sec:preliminaries}, we use the shorthand $g.D$ to refer to the set of 
values of attribute $D$ for a group $g$ when defining constraints of this type.

As indicated in \autoref{tab:constraints}, diverse constraints can be defined 
on the instance-level, relating to attribute values, associated roles, and 
duration, such as \emph{the total cost of an instance is at most 500\$} 
and \emph{the average duration of group instances must be at most 1 hour}.
As shown in the table's last row, also looser constraints may be expressed, such as ones that only need to hold for 95\% of the respective group instances.
In fact, as for class-based constraints, \approach supports all constraints of which satisfaction can be checked for an individual group $g$.
  
 \mypartwo{Satisfaction} 
We write $R_I \subseteq \userconstraints$ for the instance-based constraints. 
Contrary to the other categories, these constraints must be explicitly checked 
against the event log $L$, specifically for each group $g \in G$ and each 
instance of $g$ in the traces of~$L$.

Formally, we first define a function $\inst: \mathcal{E}^* \times 2^\mathcal{C} 
\rightarrow 2^{\mathcal{E}^*}$, which returns all instances of a group in a 
given trace. The operationalization  of $\inst$ is straightforward for simple 
cases: An instance of group $g$ is  the projection of the event classes 
of $g$ over a trace $\sigma$. In $\sigma_1$, $\sigma_2$, and 
$\sigma_3$ of our running example, exactly one instance of each group occurs 
per trace and, e.g., $\inst(\sigma_1, \gclrkone) = \{\langle \emph{rcp}, 
\emph{ckc} \rangle \}$. However, processes often include recurring behavior, 
such as trace $\sigma_{4} =\langle$\textcolor{blue}{\myuline{rcp}}, 
\textcolor{blue}{\myuline{ckc}}, rej, \textcolor{blue}{\myuline{rcp}}, 
\textcolor{blue}{\myuline{ckt}}, acc, \textcolor{blue}{\myuline{prio}}, 
\textcolor{blue}{\myuline{inf}}, \textcolor{blue}{\myuline{arv}}$\rangle$,
in which a request is first rejected, sent back to the restart the process, and 
then accepted in the second round. 
Here, to detect multiple instances of a group, we instantiate function $\inst$ 
based on an existing technique~\cite{vanderaa2021detecting} that recognizes 
when a trace contains recurring behavior and splits the (projected)  sequence 
accordingly. For the above trace, this yields $\inst(\sigma_4, \gclrkone) = 
\{\langle \emph{rcp}, \emph{ckc} \rangle,  \langle \emph{rcp}, \emph{ckt} 
\rangle \}$.
Note that $\inst$ can also be used to enforce cardinality constraints, e.g., if a user desires that each group instance should contain at least 2 events of a particular event class.


Given the function $\inst$, a constraint $r \in R_I$ is satisfied if for each 
group $g \in \finalgrouping$, $r$ holds for each instance $\xi \in 
\inst(\sigma, g)$, for each $\sigma \in L$. Note that constraints are vacuously 
satisfied for traces that do not include an instance of a particular group, 
i.e., where $\inst(\sigma, g) = \emptyset$. Therefore, for instance-based 
constraints, \holds~is checked for each $g \in \finalgrouping$ as $\forall\ r 
\in 
\userconstraints_I,\forall\ \sigma_i \in \eventlog, \forall\ \xi \in 
\inst(\sigma_i, g):  r(\xi) = \true$.
For looser constraints, e.g., ones that should for 95\% of the group instances, predicate satisfaction is adapted accordingly.

\mypartwo{Monotonicity}
As is the case for class-based ones, instance-based constraints are monotonic when they specify a minimum requirement to be met, e.g., each instance should take \emph{at least} one hour, and anti-monotonic when they specify something that may not be exceeded, e.g., each  instance may take \emph{at most} one hour. 
However, constraints in $R_I$ may also be based on aggregations that behave in a non-monotonic manner, such as constraints that consider the \emph{average} or \emph{variance} of attribute values per group instance or \emph{sums} including negative values. In these cases, adding and removing event classes from a group can result in a violated constraint to now hold or vice versa.

\subsection{Distance measure}
\label{sec:scope:distance}

To determine which event classes are suitable candidates to be grouped 
together, we employ a distance function  $\dist(\finalgrouping,\eventlog)$ that 
quantifies the relatedness of the event classes per group.
Although our work is largely independent of a specific distance function, we 
argue that log abstraction should group together event classes such that 
\begin{inparaenum}
	\item events within a group are \emph{cohesive}, i.e., the events belonging 
	to a single group instance occur close to each other, meaning there are few 
	interspersed events from other instances;
	\item events within a group are \emph{correlated}, i.e., the events 
	belonging to a single group typically occur together in the same trace and 
	group instance;
	\item larger groups are favored over \emph{unary groups}, i.e., the 
	grouping $\finalgrouping$ actually results in an abstraction.
\end{inparaenum}
To capture these three aspects, we propose the following distance function for 
an individual group $g$ and a log $L$:

\vspace{-1em}
\begin{equation}
	\label{eq:distance}
	\dist(g, L) = \hspace{-1.6em}\sum\limits_{\xi \in \inst(L, g)}\hspace{-0.5em}\frac{\frac{\interrupts(\xi)}{|\xi|} + \frac{\missing(\xi, g)}{|g|} + \frac{1}{|g|}   }{| \inst(L, g)|} 
\end{equation}

The first summand in the numerator of \autoref{eq:distance} considers cohesion. 
Here, $\interrupts(\xi)$ counts how many events from other instances are 
interspersed between the first and last events of a given group instance $\xi$. 
As such, this penalizes groups of events that are often \emph{interrupted} by 
others, 
e.g., in a trace $\langle a, b, c, d, e \rangle$, grouping $a$ and $e$ together is unfavorable, since the instance $\langle a, e \rangle$ has three interspersed events.
In \autoref{eq:distance}, the number of interruptions is considered relative to 
the length of $\xi$.
The second summand in the numerator of \autoref{eq:distance} quantifies the 
degree of completeness of $\xi$ with respect to $g$, thus capturing the 
correlation between the events in $g$. Here, $\missing(\xi, g)$ returns how 
many event classes from $g$ are missing from its instance $\xi$, which is then 
offset against the total number of classes $|g|$. 
Finally, since groups with a single event class have perfect cohesion and correlation by default, we include $\frac{1}{|g|}$ to ensure that larger groups with the same cohesion and correlation are favored, thus avoiding unary groups when possible. 


Finally, to quantify the entire distance of a grouping $G$, we sum up the distance values of all groups in $G$, resulting in the following function that will be minimized in our approach:
\begin{equation}
\label{eq:distance_overall}
\dist(\finalgrouping, \eventlog) = \sum_{g \in \finalgrouping}\dist(g, \eventlog)
\end{equation}

%% file: sections/approach.tex

Next, we describe how \approach achieves the goal of finding an optimal event 
grouping, given the distance function and constraints defined above.
\autoref{sec:approach:overview} provides a high-level overview, while 
\autoref{sec:approach:step1} to \autoref{sec:approach:step3} outline the 
algorithmic details. 

\subsection{Approach Overview}
\label{sec:approach:overview}

\begin{figure}[htbp]
	\centering
	\includegraphics[width=\linewidth]{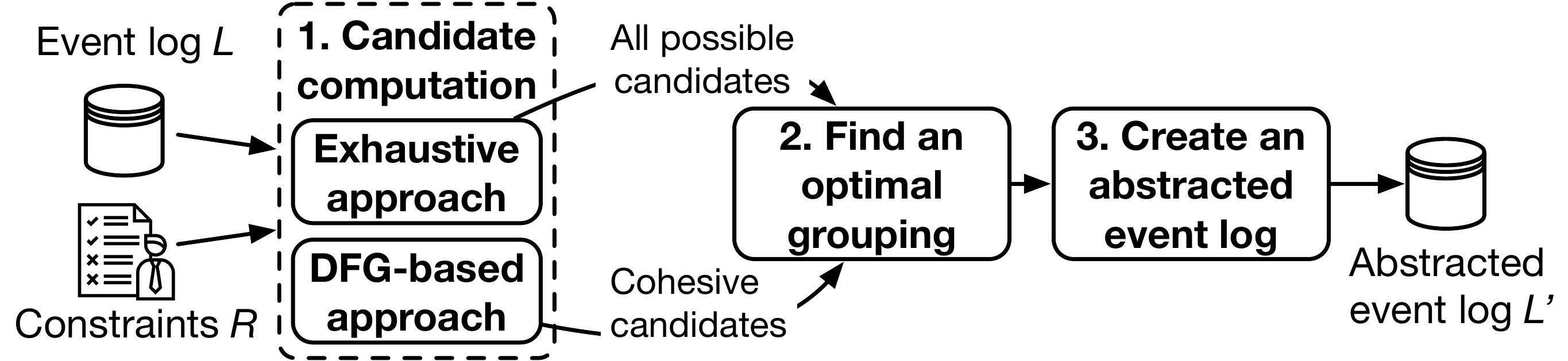}
	\caption{The main steps of \approach.}
	\label{fig:approach}
\end{figure}

As shown in \autoref{fig:approach}, \approach takes an event log \eventlog  and 
a set of user-defined constraints \userconstraints as input.
Then, \approach applies three main steps in order to obtain an abstracted log $L'$.

In Step~1, \approach computes a set of candidate groups $\allcandidates$, i.e., groups of event classes that adhere to the constraints in \userconstraints.
As depicted in \autoref{fig:approach}, we propose two instantiations for this step: an exhaustive instantiation and an efficient \dfg-based one.
The \emph{exhaustive} instantiation yields a set of candidates that is guaranteed to be complete and, thus, assures that it can afterwards be used to establish an optimal grouping if it exists.

However, 
this approach may be intractable in practice. Therefore, we also propose a 
\textit{DFG-based} alternative, which only retrieves cohesive candidates, i.e., 
candidates likely to be part of an optimal grouping. By exploiting the 
process-oriented nature of the input data, cohesive candidates are
identified efficiently. 
Any solution obtained using 
this instantiation is still guaranteed to satisfy the constraints in 
\userconstraints, yet may have a sub-optimal distance score.

In Step~2, \approach uses the identified set of candidates in order to find a single optimal grouping \finalgrouping that minimizes the distance function \dist, while ensuring that all constraints are met and each event class in $C_L$ is assigned to exactly one group in $\finalgrouping$. 
To achieve this, we formulate this task as a mixed-integer programming (MIP) problem.

Finally, having obtained a grouping $\finalgrouping$, Step~3 
abstracts event log \eventlog by replacing the events in a trace with 
activities based on groups defined in $\finalgrouping$, yielding an abstracted 
log \abstractedlog.

\subsection{Step 1: Computation of candidate groups}
\label{sec:approach:step1}

In this step, \approach computes a set of candidate groups of event classes, 
$\allcandidates$, i.e., subsets of $\eventclassesinlog$ that adhere to 
constraints in \userconstraints.  
As described in \autoref{sec:approach:overview}, we propose an exhaustive and a \dfg-based instantiation for this step:

\mypar{Exhaustive candidate computation}
To obtain a complete set of candidate groups, in principle, every combination 
of event classes, i.e., every subset of $\eventclassesinlog$, needs to be 
checked against the constraints in $\userconstraints$. However, we are able to 
considerably reduce this search space by (for the moment) only looking for 
groups $g \subseteq C_L$ that actually co-occur in at least one trace in the 
log (referred to as \emph{group co-occurrence}) and by considering the 
monotonicity of the constraints in \userconstraints. Then, candidate groups of 
increasing size can be identified iteratively, as outlined in 
\autoref{alg:exhaustivesearch}.

	\begin{algorithm}[!htb]
	\caption{Exhaustive candidate computation}
	\label{alg:exhaustivesearch}
	\small 
	\hphantom{t} \textbf{Input} Event log \eventlog, user constraints \userconstraints \\
	\hphantom{t} \textbf{Output} Set of candidate groups $\allcandidates$ 
	\begin{algorithmic}[1]
		\State \emph{mode} $\gets$ \texttt{setCheckingMode}(\userconstraints) \label{line:modechecking} 
		\State {\emph{toCheck} $\gets \{\ \{c\} $ for $c \in \eventclassesinlog\}$}\label{line:singletons} \Comment{Set the first groups to check}
		\While{\emph{toCheck} $ \neq \emptyset$ }
	
		\If{\emph{mode} = \emph{monotonic}}
		\State{$ \allcandidates_{new} \gets \{g \in$ \emph{toCheck} if \label{line:monocheck} 
		$\exists g' \in \allcandidates: g' \subset g $ \newline \hphantom{xxxxxxxxxxxxxxxxxxxxxxxxx} 
		$\lor\ \holds(g,\eventlog,\userconstraints)$ }	
	
		\Else
		\State{$ \allcandidates_{new} \gets \{g \in$ \emph{toCheck} if $  \holds(g,\eventlog,\userconstraints)\} $} \label{line:regcheck}
		\EndIf
		\State{$\allcandidates \gets \allcandidates~\cup  \mathcal{G}_{new}$ } \label{line:amout}
		\If{\emph{mode} = \emph{anti-monotonic}}
		\State{\emph{toCheck} $\gets$ \texttt{expandGroups}($\allcandidates_{new}$)} \label{line:expandAM}
		\Else
		\State{\emph{toCheck} $\gets$ \texttt{expandGroups}(\emph{toCheck})} \label{line:expandreg}
		\EndIf
		\State{\emph{toCheck} $\gets \{g \in$ \emph{toCheck}  if $ \texttt{occurs}(g, L)\} $}  \label{line:occurs}
		\EndWhile
		
		\State \textbf{return} $\allcandidates$
	\end{algorithmic}
\end{algorithm}

\mypartwo{Initialization} 
We first set the constraint-checking \emph{mode} that shall be applied (line~\ref{line:modechecking}) based on the monotonicity of the constraints in \userconstraints. Specifically, \emph{mode} is set to \emph{anti-monotonic} if \userconstraints contains at least one such constraint, to \emph{monotonic} if all constraints in $\userconstraints\setminus\userconstraints_G$ are monotonic (i.e., all constraints that must be checked \emph{per group}), and otherwise to \emph{non-monotonic}.

Using this constraint-checking mode, we employ two pruning strategies.
First, consider a group $g_1 \subset C_L$ and a constraint set 
$\userconstraints$ in which  all constraints 
$\userconstraints\setminus\userconstraints_G$ are  monotonic. 
If $\holds(g_1, 
\userconstraints, L) = \true$, any supergroup $g'_1 \supseteq g_1$ will also 
adhere to the constraints, since adding more event classes to $g_1$ will 
never lead to a violation of a monotonic constraint. 
Therefore, in the \emph{monotonic} mode, we can avoid the costs of constraint validation for $g'_1$.
Second, consider a group $g_2\subset C_L$, known to violate any anti-monotonic constraint in $\userconstraints$, i.e, 
$\holds(g_2, \userconstraints, L) = \false$. 
Then we also know that no supergroup $g'_2 
\supseteq g_2$ can adhere to \userconstraints, as expanding a group can 
never resolve violations of anti-monotonic constraints. Thus, in the 
\emph{anti-monotonic} mode, all supergroups of $g_2$ can be skipped.

With \emph{mode} set, the algorithm then adds all event classes of $\eventclassesinlog$ as singleton groups to the set of potential candidates, \emph{toCheck}, which shall be checked in the first iteration (line~\ref{line:singletons}).

\mypartwo{Candidate assessment} 
In each iteration, \autoref{alg:exhaustivesearch} first 
establishes a set $ \allcandidates_{new}$, which contains all groups in \emph{toCheck} that adhere to the constraints in \userconstraints. When validating a group, we check constraints in $\userconstraints_C$ before ones in $\userconstraints_I$, since the former do not require a pass over the event log, thus, minimizing the validation cost per candidate.
In the \emph{monotonic} mode, the algorithm employs the first pruning strategy by directly adding any  group $g$, for which there is a $g' \subset g$ already in $\allcandidates$, given that we then know that the monotonic constraints will be satisfied for $g$ as well (line~\ref{line:monocheck}). 
For other groups and for the other two modes, we need to check $\holds(g, L, R)$ for each group $g \in \mathit{toCheck}$ (lines~\ref{line:monocheck} and~\ref{line:regcheck}).
Having established $ \allcandidates_{new}$, the new candidates are added to the total set $\allcandidates$ (line~\ref{line:amout}).

\mypartwo{Group expansion}
Next, the algorithm repopulates \emph{toCheck} with larger groups that shall be assessed in the next iteration. In the \emph{anti-monotonic} mode, using the second pruning strategy, the algorithm only needs to expand groups that are known to adhere to all anti-monotonic constraints in $R$. Therefore, in this case, we only expand the groups in $\allcandidates_{new}$ (line~\ref{line:expandAM}). This expansion involves the creation of new groups that consist of a group $g \in  \allcandidates_{new}$ with an additional event class from $C_L$. Naturally, the \emph{anti-monotonic} mode avoids the creation of groups that contain subgroups that are already known to violate $\userconstraints$. For the \emph{monotonic} and \emph{non-monotonic} mode, we also need to expand groups that currently violate the constraints, since their supergroups may still lead to constraint satisfaction. Therefore, these modes expand all groups in \emph{toCheck} (line~\ref{line:expandreg}).
Afterwards, we only retain those groups in \emph{toCheck}$(g, L)$ that actually occur in the event log, by checking if there is at least one trace in $L$ that contains events corresponding to all event classes in $g$ (line \ref{line:occurs}).

 \mypartwo{Termination} The algorithm stops if there are no new candidates to be checked, returning the set of all candidates, $\allcandidates$.

\mypartwo{Computational complexity}
While \autoref{alg:exhaustivesearch} is guaranteed to yield a complete set of candidates, its time complexity is exponential with respect to the number of event classes in the event log, i.e., $2^{|\eventclassesinlog|}$.
In the worst case, each of the subsets of $\eventclassesinlog$ must be analyzed against the entire log, where primarily the number of traces is important, since each group must be separately checked against all traces. 
Given that each checked group may become a  candidate, the algorithm's space complexity is also bounded by $2^{|\eventclassesinlog|}$.
Hence, this exhaustive approach can quickly become infeasible.

\mypar{\dfg-based candidate computation}
In the light of the runtime complexity of the exhaustive approach, we also 
propose a \dfg-based approach to compute candidate groups. It exploits 
behavioral regularities in event logs in order to efficiently derive a set of 
\emph{cohesive candidate groups}. 

\mypartwo{Intuition}
Log abstraction aims to find cohesive groups of event classes and, therefore, 
is more likely to group together event classes that occur close to each other.
In our running example, even though the \emph{request receipt} 
(\emph{rcp}) and \emph{archive request} (\emph{arv}) event classes meet the 
constraint (both are performed by a \emph{clerk}), it is 
unlikely that they will end up in the same activity in an 
optimal grouping $G$, since \emph{rcp} occurs at the start of each trace and 
\emph{arv} at the end.

We exploit this characteristic of optimal groupings by identifying only 
candidates that occur near each other. 
This is achieved by establishing a \dfg of the event log  and traversing this 
graph to find highly cohesive candidates groups. 
Since this traversal again iteratively increases the candidate size, we can 
still apply the aforementioned pruning strategies.

\begin{figure}[!htb]
	\centering
	\includegraphics[width=0.8\linewidth]{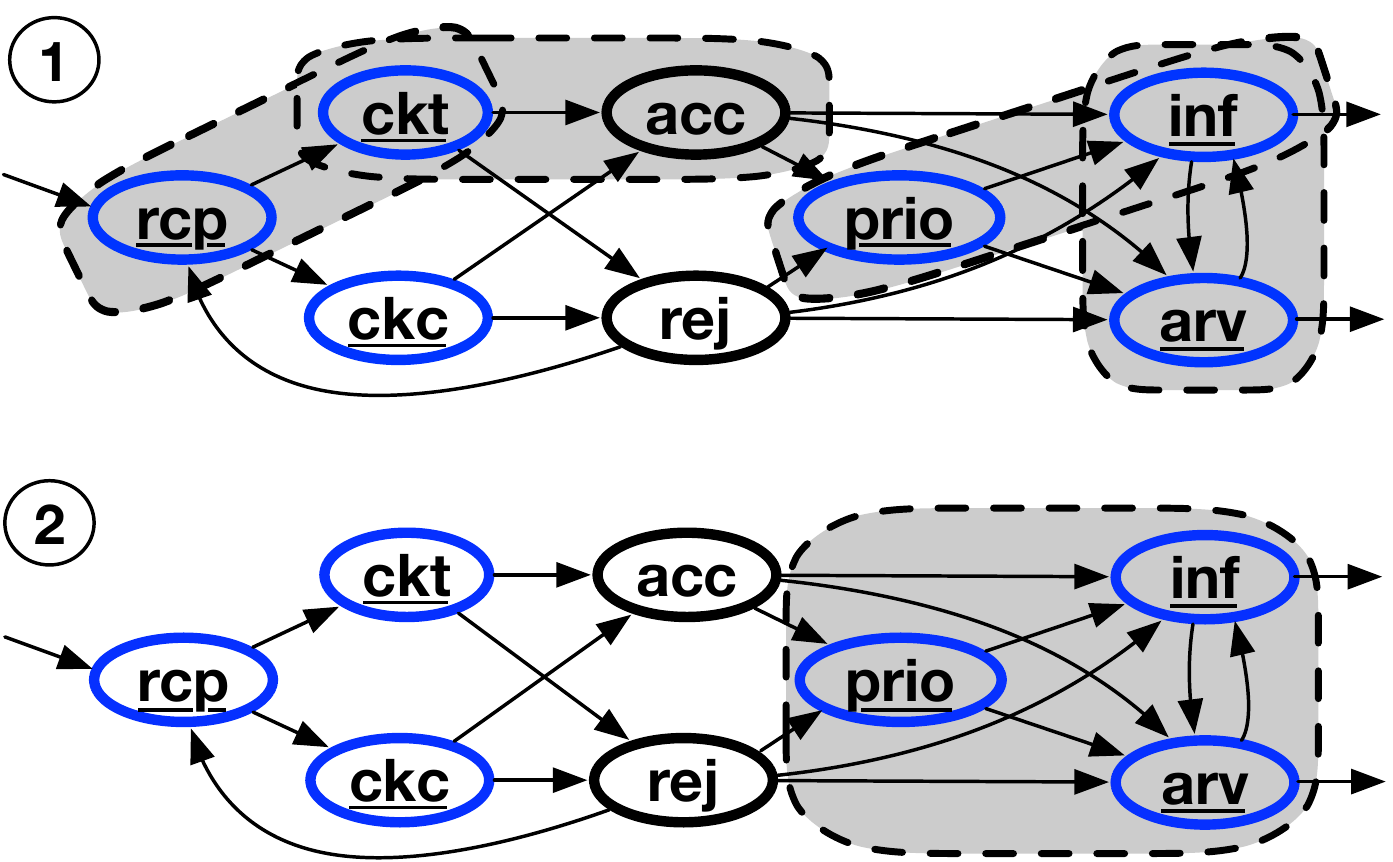}
	\caption{\dfg-based candidate computation (Iterations 1 \& 2).}
	\label{fig:dfgillustration}
\end{figure} 

This idea is illustrated in \autoref{fig:dfgillustration}, which visualizes (parts of) two iterations for the running example, highlighting candidate groups that are checked. 
Iteration 1 involves the assessment of paths of length two, consisting of 
connected event classes.
This identifies, e.g., the candidate paths [\emph{prio,inf}],  
[\emph{prio,arv}], and  [\emph{inf,arv}], which all adhere to the constraint, 
whereas, e.g., [\emph{acc, inf}] is recognized as a violating 
path, since \emph{acc} and \emph{inf} are performed by different roles.
Given their distance from each other in the \dfg, this iteration avoids 
checking groups such as \{\emph{rcp, arv}\} and \{\emph{ckt, inf}\}.
In the next iteration, since the running example deals with an 
\emph{anti-monotonic} constraint, we concatenate pairs of constraint-adhering 
paths to obtain candidate paths (i.e., groups) of length three, as shown for 
[\emph{prio, inf, arv}] in \autoref{fig:dfgillustration}.



\smallskip
The \dfg-based approach works as described in \autoref{alg:efficientmining}. 
Next to an event log $L$ and a constraint set $\userconstraints$, it takes as input a parameter $k$, defining the beam-search width. 

\mypartwo{Initialization}
The algorithm starts by again setting the constraint-checking mode (line~\ref{line2:mode}), before establishing the log's \dfg (line~\ref{line2:dfg}), as defined in \autoref{sec:preliminaries}.
Then, for every node $n$ in the DFG (i.e., for every event class), we add the trivial path $\langle n \rangle $ to the set of candidates to check in the first iteration (line~\ref{line2:singletons}). 

\begin{algorithm}[!tb]
	\caption{DFG-based candidate computation}
	\label{alg:efficientmining}
	\small
	\hphantom{t} \textbf{Input} event log \eventlog, user constraints \userconstraints, pruning parameter $k$ \\
	\hphantom{t} \textbf{Output} Set of candidate groups $\allcandidates$ 
	\begin{algorithmic}[1]
		\State \emph{mode} $\gets$ \texttt{setCheckingMode}(\userconstraints) \label{line2:mode}
		\State $\mathit{DFG} \gets$ \texttt{computeDFG($L$)} \label{line2:dfg}
		\State {\emph{toCheck} $\gets \{ \langle n \rangle$ for $n \in \mathit{DFG}.\texttt{nodes} \}$}\label{line2:singletons} \Comment{First paths to check}
		\While{\emph{toCheck} $\neq \emptyset$} 
		\State \emph{sortedPaths} $\gets \texttt{sort}(\textit{toCheck}, \dist)$ \label{line2:sortpaths} \Comment{Lowest \dist first}	
		\State {$\emph{toExpand} \gets \emptyset$} 
		\State{$i \gets 0$}
		\While{$i < \min(|\emph{sortedPaths}|, k)$}	 \label{line2:bfsstart}		
			\State $p \gets \emph{sortedPaths}[i]$ \Comment{Get next path}
			\State $g \gets \texttt{nodes}(p)$ \Comment{Derive nodes, i.e., event classes of $p$}
			\If{\emph{mode} = \emph{monotonic}}
				\If{$\exists g' \in \allcandidates: g' \subset g \lor\ \holds(g,\eventlog,\userconstraints)$ } \label{line2:checkmono}
							\State {$\allcandidates \gets \allcandidates \cup \{g\}$}  \label{line2:addtoresultm}
				\EndIf
			\State $toExpand \gets toExpand \cup \{ p \}$
			\ElsIf{$\holds(g, L, R)$ \label{line:addtoresult}}
			\State {$\allcandidates \gets \allcandidates \cup \{g\}$} \label{line2:addtoresult}
			\State $toExpand \gets toExpand \cup \{ p \}$
			\ElsIf{$mode \neq $ \emph{anti-monotonic}} \label{line2:checkantimono}
			\State{$toExpand \gets toExpand \cup \{ p \}$}\label{line2:antimono}
			\EndIf
			\State{$i \gets i +1$}
		\EndWhile \label{line:beamcheckend}
		\State {\emph{toCheck} $\gets \emptyset$} \label{line:clearb} \Comment{Start computing new paths to check}
		\For{$p = \langle p_{0}, \ldots, p_{m} \rangle \in \emph{toExpand} $ }\label{line2:mergestuffstart} 
			\For{$(p_m, \emph{succ}) \in \mathit{DFG}.\texttt{outgoingEdges}(p_m)$  }
				\If{$\emph{succ} \notin \texttt{nodes}(p)$}
					\State $\emph{toCheck} \gets \emph{toCheck} \cup \{\ \langle p_{0}, \ldots, p_{m}, \emph{succ} \rangle~\}$ \label{line2:tocheck1}
				\EndIf
			\EndFor
			\For{$(\emph{pred},p_0 ) \in \mathit{DFG}.\texttt{incomingEdges}(p_0)$  }
			\If{$\emph{pred} \notin \texttt{nodes}(p)$}
			\State $\emph{toCheck} \gets \emph{toCheck} \cup \{\ \langle \emph{pred}, p_{0}, \ldots, p_{m} \rangle~\}$  \label{line2:tocheck2}
			\EndIf
			\EndFor
		\EndFor  \label{line2:mergefornextend}
		\State{\emph{toCheck} $\gets \{p \in$ \emph{toCheck}  if $ \texttt{occurs}(\texttt{nodes}(p), \eventlog)\} $}  \label{line2:occursdfg}
		\EndWhile
		\State \textbf{return} $\allcandidates$
	\end{algorithmic}
\end{algorithm}

\mypartwo{Candidate assessment}
In principle we could assess for each path $p \in \emph{toCheck}$ if $p$'s nodes form a proper candidate group, as we do in the exhaustive approach. However, we here recognize that in event logs with a lot of variability, the number of paths to check will still be considerable.
Hence, we allow for a further pruning of the search space by incorporating a \emph{beam-search}~\cite{Wilt2010} component in the algorithm. 
In this beam search, we only keep the $k$ most promising candidates (i.e., the \emph{beam}) in each iteration of the algorithm. 

To do this, each iteration starts by sorting the candidate paths in \emph{toCheck}, giving priority to paths of which the nodes have the lowest distance to each other, according to  $\dist(\texttt{nodes(p)}, L)$  (line \ref{line2:sortpaths}).
Then, the algorithm picks candidates from \emph{sortedPaths} as long as there are candidates to pick and the beam width $k$ has not been reached (line \ref{line2:bfsstart}).
Each group $g$, defined by the nodes in a path (i.e., $g = \texttt{nodes}(p)$), is then checked for constraint satisfaction. As for the exhaustive approach, we check constraints in $\userconstraints_C$ before ones in $\userconstraints_I$ minimizing validation cost per candidate.

Here, we employ the same pruning strategies for \emph{monotonic} and \emph{anti-monotonic} constraint-checking modes as done for the exhaustive approach.
Therefore, in the \emph{monotonic} mode, a group $g$ can be directly added to the set of candidates $\allcandidates$ if we have already seen a subset $g' \subset g$ that adheres to the constraints (lines~\ref{line2:checkmono}--\ref{line2:addtoresultm}), whereas in the \emph{anti-monotonic} mode, we no longer expand paths that violate the constraints (lines~\ref{line2:checkantimono}--\ref{line2:antimono}).

\mypartwo{Path expansion}
The candidates for the next iteration are created by expanding paths in \emph{toExpand} with either a predecessor of their first or a successor of their last node  (lines \ref{line2:mergestuffstart}--\ref{line2:mergefornextend}). 
Again, we then only retain those paths in \emph{toCheck}, whose groups $g$ actually occur in the event log (line \ref{line2:occursdfg}). 

\mypartwo{Termination} The algorithm stops if no candidates are left, i.e., \emph{toCheck} is empty and the set $\allcandidates$ is returned.

\mypartwo{Computational complexity}  
	The \dfg-based approach is considerably more efficient than the exhaustive one. In each iteration, the approach expands up to $k$ groups, each into up to $|\eventclassesinlog|-1$ new candidates. As such, given the maximum of $|\eventclassesinlog|$ iterations, the worst-case time and space complexity is $k * |\eventclassesinlog|^2$. Moreover, this worst case only occurs if the \dfg is a complete digraph and no constraints are imposed.

\mypar{Dealing with exclusion}
Generally, it is undesirable to group exclusive event classes together, since such classes never occur in the same trace. This is why we have so far omitted these from consideration, by ensuring that  \texttt{occurs}$(g, L)$ holds for every candidate group $g \in \allcandidates$.
Yet, when exclusive event classes (or groups) are proper alternatives to each other, we make an exception for this. In these cases, grouping them together will result in a further complexity reduction of the event log, while not affecting its expressiveness.

\begin{figure}[!htb]
	\centering
	\includegraphics[width=0.8\linewidth]{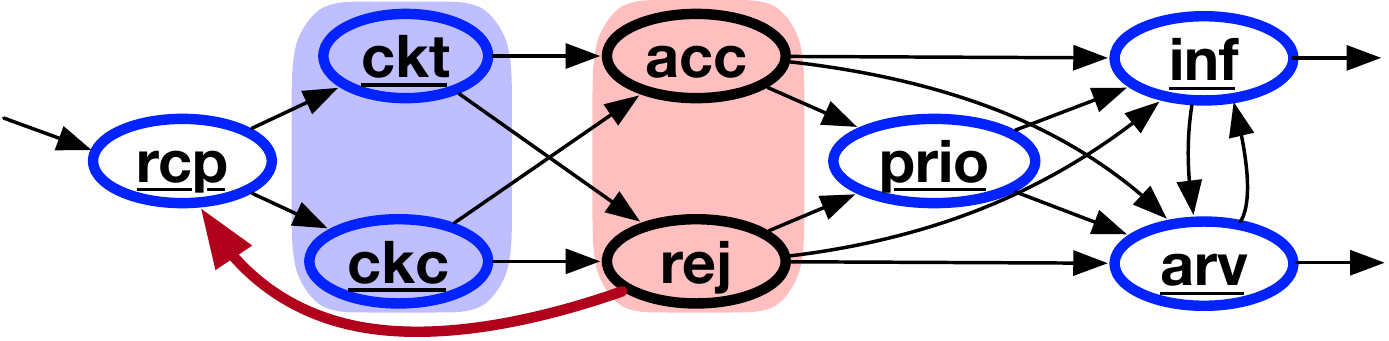}
	\caption{\dfg of the running example highlighting proper behavioral alternatives (blue) and exclusive event classes, which are no behavioral alternatives (red).}
	\label{fig:exclusive}
\end{figure}

\mypartwo{Intuition}
To illustrate this, reconsider the running example, which contains two sets of 
exclusive event classes, \{\emph{ckc}, \emph{ckt}\}, corresponding to two ways 
in which a request can be checked, and \{\emph{acc}, \emph{rej}\}, 
corresponding to acceptance and rejection of a request. By considering 
\autoref{fig:exclusive}, we see that the former two event classes are proper 
behavioral alternatives: both \emph{ckc} and \emph{ckt} are preceded and 
followed by the exact same sets of event classes.As such, 
behavioral alternatives can be defined as groups of event classes that have 
identical \emph{pre-} and \emph{postsets} in the \dfg. Merging them will thus 
not lead to a loss of behavioral information.
By contrast, \emph{acc} and \emph{rej} do not represent proper alternatives to each other, since their postsets differ. Particularly, while after acceptance the process always moves forward to one of the event classes in \{\emph{prio}, \emph{inf}, \emph{arv}\}, a rejection may also result in a loop back to the start of the process (\emph{rcp}). Therefore, if these exclusive classes were merged, we would obscure the fact that there are two different possibilities here. 

\mypartwo{Candidate identification}
We employ \autoref{alg:exclusive} to determine if previously identified candidate groups in $\allcandidates$, with excluding event classes, can be merged to obtain additional candidates.

The algorithm establishes a set \emph{equivGroups} consisting of candidate groups that share the same pre- and postset (line~\ref{line3:equalprepost}). Then, a stack is created consisting of all pairs of groups in this set (lines~\ref{line3:stackcreate}--\ref{line3:stackpopulate}). For each pair $(g_i, g_j)$ in this stack, we assess if $g_i$ and $g_j$ are indeed exclusive to each other and if their merged group, $g_{ij}$, still adheres to the user constraints (line~\ref{line3:check}). 
Both conditions can be efficiently checked. The former by ensuring that there are no edges from nodes in $g_i$ to nodes in $g_j$ or vice versa, while for the latter only adherence to class-based constraints ($R_C$) needs to be assessed, given that instance-based constraints cannot be (newly) violated when merging exclusive groups, thus avoiding a pass over the event log $L$.

If $g_{ij}$ is indeed a proper, new candidate, we next determine if this group can also be combined together with its preset, postset, or with both, to create more candidates (lines~\ref{line3:mergeprepoststart}--\ref{line3:mergeprepostend}). For instance, having identified \{\emph{ckt, ckc}\} as a new candidate group for the running example, we would this way recognize that this new group together with its preset (event class \emph{rcp}) also forms a proper candidate group: \{\emph{rcp, ckt, ckc}\}, since both \{\emph{rcp, ckt}\} and \{\emph{rcp, ckc}\} were also already part of $\allcandidates$.

After establishing these new candidates, the algorithm adds any new pair $(g_{ij}, g_k)$ to the stack, so that also iteratively larger candidates, consisting of three or more exclusive groups, can be identified (lines~\ref{line3:expansionstart}--\ref{line3:expansionend}). The algorithm terminates when all relevant pairs have been assessed, returning the updated set $\allcandidates$ as the final output of Step~1 of the approach.

	\begin{algorithm}[!htb]
	\caption{Find exclusive candidate groups}
	\label{alg:exclusive}
	\small 
	\hphantom{t} \textbf{Input} Event log \eventlog,  user constraints \userconstraints, current candidate groups $\allcandidates$\\
	\hphantom{t} \textbf{Output} Extended set of candidate groups $\allcandidates$ 
	\begin{algorithmic}[1]
		\State $\mathit{DFG} \gets$ \texttt{computeDFG($L$)} \label{line3:dfg}
		\State $\emph{seenGroups} \gets \emptyset$
		\For{$g \in \mathcal{G} \setminus \emph{seenGroups}$}
			
			\State \emph{equivGroups} $\gets\mathit{DFG}$.\texttt{equalPrePost}($g$) $\cup\ \{g\}$ \label{line3:equalprepost}
			\State \emph{pairsToCheck} $\gets \texttt{new Stack}()$ \label{line3:stackcreate}
			\For{$(g_i, g_j) \in \emph{equivGroups} \times \emph{equivGroups}$} 
					\State \emph{pairsToCheck}.\texttt{push}$\left((g_i, g_j) \right)$ \label{line3:stackpopulate}
			\EndFor
			
			\While{$\neg \emph{pairsToCheck}.\texttt{isEmpty}()$}
				\State $(g_i, g_j) \gets \emph{pairsToCheck}.\texttt{pop}()$
				\State{$g_{ij} \gets g_i \cup g_j$ } \Comment{Merge into single group}
				\If{\texttt{exclusive}($g_i, g_j) \land \texttt{holds}(g_{ij}, L, R_C)$} \label{line3:check}
				\State{$\allcandidates \gets \allcandidates \cup \{g_{ij} \}$} \Comment{New candidate found}
				\State (\emph{pre, post}) $\gets(\mathit{DFG}$.\texttt{pre}($g_i$), $\mathit{DFG}$.\texttt{post}($g_i$)) \label{line3:mergeprepoststart}
				\If{\emph{pre} $\cup$ \emph{post} $\cup\ g_i \in \allcandidates~\land$ \emph{pre} $\cup$ \emph{post} $\cup\ g_j \in \allcandidates$}  
				\State{$\allcandidates \gets \allcandidates\cup \{\emph{pre} \cup \emph{post} \cup g_{ij} \}$}
				\ElsIf{\emph{pre} $\cup\ g_i \in \allcandidates~\land$ \emph{pre} $\cup\ g_j \in \allcandidates$}  
				\State{$\allcandidates \gets \allcandidates\cup \{$\emph{pre} $\cup~g_{ij}\}$}
				\ElsIf{\emph{post} $\cup~g_i \in \allcandidates~\land~$\emph{post} $\cup~g_j \in \allcandidates$}
				\State{$\allcandidates \gets \allcandidates\cup \{$\emph{post} $\cup~g_{ij}\}$} \label{line3:mergeprepostend}
				\EndIf
				
				\For{$g_k \in \emph{equivGroups} \setminus \{g_i, g_j\}$ } \label{line3:expansionstart}
						\State \emph{pairsToCheck}.\texttt{push}$\left((g_{ij}, g_k) \right)$\label{line3:expansionend}
				\EndFor
				\State \emph{equivGroups} $\gets \emph{equivGroups} \cup \{ g_{ij} \}$
				\EndIf 
			\EndWhile
			\State \emph{seenGroups} $\gets \emph{seenGroups} \cup \emph{equivGroups}$
		\EndFor 
		\State \textbf{return} $\allcandidates$

	\end{algorithmic}
\end{algorithm}

	\mypartwo{Computational complexity} 
	\autoref{alg:exclusive} has linear complexity with respect to 
	$|\allcandidates|$, i.e., the number of candidate groups stemming from the 
	previous step.
	As such, its worst-case time and space complexity is $2^{|\eventclassesinlog|}$ when previously using the exhaustive approach and $k * |\eventclassesinlog|^2$ for the \dfg-based one.

\subsection{Step 2: Finding an optimal grouping}
\label{sec:approach:step2}

Having established candidate groups $\allcandidates$, we set out to find an optimal grouping $\finalgrouping \subseteq \allcandidates$ based on these candidates, which is a set of disjoint groups  that covers all event classes, while minimizing the overall distance.
We formulate this task as a MIP problem, which 
can be tackled using standard solvers. 


\begin{figure}[!htb]
	\centering
	\includegraphics[width=\linewidth]{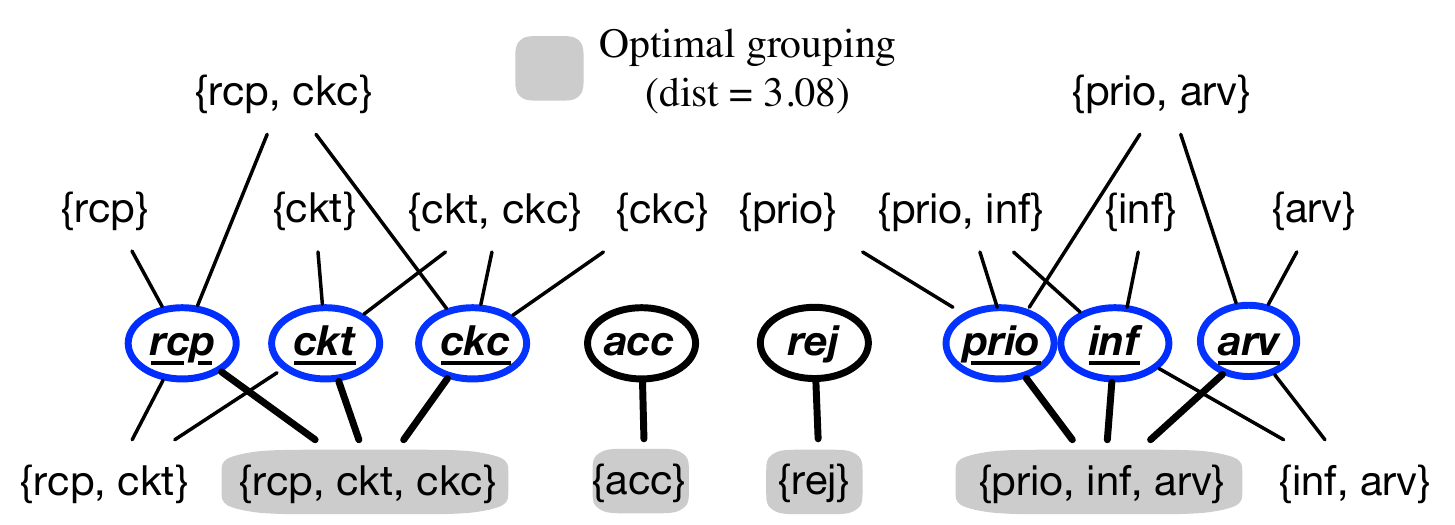}
	\caption{Optimal grouping of the running example given all candidates computed in step 1 using the \dfg-based approach.}
	\label{fig:exactcover}
\end{figure} 

Central to this formulation is a bipartite graph $(\allcandidates, C_L, E)$, which connects each candidate group to the event classes it covers, i.e., it contains an edge $(g_i, c_j) \in E$ if $c_j \in g_i$. \autoref{fig:exactcover} visualizes this for the running example, in which the circled nodes in the middle indicate event classes in $C_L$, the sets indicate the candidate groups $\allcandidates$, and the edges their coverage relation. The grayed sets highlight the optimal grouping in this case, which is an exact cover  because every event-class node is connected to exactly one of the selected groups.

Given this bipartite graph $(\allcandidates, C_L, E)$, we formalize a MIP problem with two decision variables:

\begin{compactitem}
		\item $ s e l e c t e d_{g_i} \in \{0,1\}$: 1 if $g_i \in \allcandidates$ is selected, else 0;
		\item $c o v e r e d_{c_j} \in \{0,1\}$: 1 if $c_j \in \eventclassesinlog$ is covered, else 0.
\end{compactitem}
Then, we seek to minimize the distance of the selected groups in the objective function:
\begin{equation*}
\label{eq:objective}
\text{minimize}\  \displaystyle\sum\limits_{g_i \in \mathcal{G}} \dist(g_i) * selected_{g_{i}}
\end{equation*}
This objective function is subject to two constraints:
\begin{equation}
	\label{eq:coverage}
	\displaystyle\sum\limits_{j=1}^{|\eventclassesinlog|}    c o v e r e d_{c_j} = |\eventclassesinlog|
\end{equation}

\begin{equation}
\label{eq:exactcoverage}
 \displaystyle\sum\limits_{(g_i,c_j) \in E}  s e l e c t e d_{g_i}  = c o v e r e d_{c_j}, \forall c_j \in \eventclassesinlog
\end{equation}

These constraints jointly express that each event class  shall be covered (\autoref{eq:coverage}), by exactly one group (\autoref{eq:exactcoverage}).
In case a user imposes grouping constraints in $R_G$, bounding the number of groups that may be selected, these are imposed by adding either or both of the following additional constraints:

\begin{equation}
\label{eq:const3}
\begin{array}{ll@{}ll}
 \displaystyle\sum\limits_{i=1}^{|\allcandidates|} ~s e l e c t e d_{g_i}  \leq x &\text{resp.}& & \displaystyle\sum\limits_{i=1}^{|\allcandidates|} ~s e l e c t e d_{g_i} \geq y
\end{array}
\end{equation}

The selected groups, i.e., $\finalgrouping = \{g_i \in \mathcal{G} : \emph{selected}_{g_i} = 1\}$, then form the obtained grouping.
 
Note that, depending on the characteristics of log $L$ and the imposed constraints $R$,  a grouping may not be found, since there is no guarantee that a feasible solution exists. In that case, \approach returns the initial log.
To allow users to then refine their constraints appropriately, \approach also indicates possible causes of the infeasibility, e.g., the affected event classes that lead to violations for constraints in $\userconstraints_C$, or the fraction of cases for which constraints in $\userconstraints_I$ are violated. If a solution is found, \approach continues with its third and final step.

	\mypartwo{Computational complexity} 
	Most MIP problems are NP-hard, although an assessment of the exact 
	complexity depends on the concrete problem and is poorly characterized by 
	input size~\cite{mipcomplexity}. 
		However, solvers like \emph{Gurobi} are often able to solve MIP 
		problems efficiently, by applying pre-solvers and heuristics.
	 Our experiments confirm this, showing that Step 2 only contributes marginally to the overall runtime of \approach.

\subsection{Step 3: Creating an abstracted event log}
\label{sec:approach:step3}

Finally, we use the grouping $\finalgrouping$ to establish abstracted versions of the traces in $L$  to obtain an abstracted log $L'$. 

For each trace $\sigma \in L$, we identify all activity instances in the trace, 
i.e., all instances of groups in $\finalgrouping$, $\mathcal{I}^\sigma = 
\bigcup_{g \in \finalgrouping} \inst(\sigma, g)$. Each activity instance, 
$\xi_i \in \mathcal{I}^\sigma$, 
corresponds to an ordered sequence of events, $ \langle e_1,...,e_k \rangle$.

Next, our approach creates an abstracted trace $\sigma'$, which reflects the 
activity instances in $\mathcal{I}^\sigma$, instead of the events of its 
original counterpart, $\sigma$. A common abstraction strategy is to let 
$\sigma'$ capture only the \emph{completion} of activity instances, by creating 
a projection of $\sigma$ that only retains the last event, $e_k$, per activity 
instance.
For instance, for trace $\sigma_1 = \langle$\textcolor{blue}{\myuline{rcp}}, 
\textcolor{blue}{\myuline{ckc}}, acc, \textcolor{blue}{\myuline{prio}}, 
\textcolor{blue}{\myuline{inf}}, \textcolor{blue}{\myuline{arv}}$\rangle $ of 
the running example, this abstraction would yield $\sigma^c_1 = 
\langle$\textcolor{blue}{\myuline{clrk1}}, {acc}, 
\textcolor{blue}{\myuline{clrk2}}$\rangle$.

Yet, this strategy may obscure information when activities are executed in an interleaving manner. For this, consider a new trace 
	$\sigma_5 =\langle$\textcolor{blue}{\myuline{rcp}}, \textcolor{blue}{\myuline{ckc}}, \textcolor{darkgray}{\myuline{prio}}, acc, \textcolor{darkgray}{\myuline{inf}}, \textcolor{darkgray}{\myuline{arv}}$\rangle $. Here, events belonging to the \emph{clrk2} group occur both before (\emph{prio}) and after (\emph{inf, arv}) the unary activity instance \emph{acc}.
When only retaining completion events, this yields the trace $\sigma^{c}_{5'} =\langle$\textcolor{blue}{\myuline{clrk1}}, acc, \textcolor{darkgray}{\myuline{clrk2}}$\rangle $, which hides the interleaving nature of the activities.
	
	Therefore, we also propose an alternative strategy, which retains both the  \emph{start (s)} and \emph{completion (c)} events per activity instance $\xi_i \in \mathcal{I}^\sigma$. This yields a trace 	$\sigma^{s+c}_{5'} =\langle\textcolor{blue}{\underline{\clrkone_s}}, \textcolor{blue}{\underline{\clrkone_c}}$, $\textcolor{darkgray}{\underline{\clrktwo_s}}$, acc, $\textcolor{darkgray}{\underline{\clrktwo_c}} \rangle $, which thus shows that activity \emph{clrk2} starts before \emph{acc} and completes afterwards.
	
	The choice for a particular strategy depends on the relevance of parallelism in a particular analysis context, given that the latter strategy also leads to longer traces, thus partially mitigating the benefits of the obtained log abstraction. 
	
	The log $L'$ that results from this last step represents the final output of \approach. It
	is an event log in which the high-level activities are guaranteed to 
	satisfy the user-defined constraints in $\userconstraints$ while providing  
	a maximal degree of abstraction.

%% file: sections/evaluation.tex

We evaluated \approach through evaluation experiments using a collection of real-world event logs.
\autoref{sec:setup} outlines the evaluation setup. \autoref{sec:realresults} reports on the  results
obtained for our approach and its configurations, whereas \autoref{sec:baselineresults} compares our work against three baselines.
Finally, \autoref{sec:casestudy} further illustrates the value of constraint-driven log abstraction through a case study.
The employed implementation, evaluation pipelines, and additional experimental results are all publicly available.\footnote{\url{https://gitlab.uni-mannheim.de/processanalytics/gecco}}

\subsection{Evaluation Setup}
\label{sec:setup}

\mypar{Implementation and environment}
We implemented our approach in Python, using \emph{PM4Py}~\cite{pm4py} for event log handling and \emph{Gurobi}~\cite{gurobi} as a solver for MIP problems.
All experiments were conducted single-threaded on an Intel Xeon 2.6 GHz processor with up to 768GB of RAM available.

\mypar{Data collection}
We use a collection of 13 publicly-available event logs.
	To be able to cover various constraints, all logs have at least one categorical event attribute, as well as timestamps used for numerical constraints.
	As shown in \autoref{table:reallogstatsdetail}, the logs vary considerably in terms of key characteristics, such as the number of event classes, traces, and variants.

\begin{table}[!htb]
	\small
	\centering
	\caption{Properties of the real-life log collection.}
	\label{table:reallogstatsdetail}
	\begin{tabular}
		{l rrrrr}
		\toprule
		\textbf{Ref} &
		$\mathbf{|\eventclassesinlog|}$ & 
		 \textbf{Traces}& 
		\textbf{Variants} & 
		 $\mathbf{|E|}$ & 
		 \textbf{Avg} $\mathbf{|\sigma|}$ \\ 
		\midrule
\cite{rtfm}     & 11 & 150,370 & 231    & 70  &   ~3.73\\
\cite{bpi2019}  & 40 & 75,928 & 3,453 & 357 & ~6.35  \\
\cite{bpi2014}  & 39 & 46,616  & 22,632 & 772 &10.01\\
\cite{bpi2017}  & 24 & 31,509  & 5,946  & 180 & 16.41\\
\cite{bpic2018} & 39 & 14,550  & 8,627& 407 & 52.48\\
\cite{bpi2012}  & 24 & 13,087  & 4,366  & 125 &20.04  \\
\cite{credit}   & 8  & 10,035  & 1      & 14  & 15.00\\
\cite{bpi2020}  & 51 & 7,065   & 1,478  & 553 &  12.25\\
\cite{bpi2013}  & 4  & 1,487   & 183    & 10  &  ~4.47\\
\cite{wabo}     & 27 & 1,434   & 116    & 99  &  ~5.98\\
\cite{sepsis}   & 16 & 1,050   & 846    & 115 &  14.49\\
\cite{bpi151}   & 70 & 902     & 295    & 124 & 24.00  \\
\cite{ccc2019}  & 29 & 20      & 20     & 164 &  69.70\\
		\bottomrule
	\end{tabular}	
	\vspace{-1.5em}
\end{table}

\mypar{Constraints}
We use ten constraint sets in our experiments, covering the various constraint types that \approach supports.
Each set includes the class-based constraint $|g| \leq 8$,  which is used to limit the number of abstraction problems that time out.
This constraint is combined with each of the sets from \autoref{tab:evalconstraints}, covering anti-monotonic ($A$), monotonic ($M$), and non-monotonic ($N$) instance-based constraints,
a grouping constraint (\emph{Gr}), as well as two sets of their combinations ($C1$ and $C2$). 
\autoref{tab:evalconstraints} also contains additional constraints ($BL1$ to $BL4$) used in baseline comparisons, described below.
By combining these constraint sets with the 13 event logs, we establish a total of 121 abstraction problems to be solved.\footnote{The class-based $BL3$ constraint can only be applied to 4 out of 13 logs, due to the absence of class-level attributes in the others.}

\begin{table}[!h] 
	\small
	\centering
	\caption{Constraints used in the experiments.}
	\label{tab:evalconstraints}
	\begin{tabular}
		{cc l@{\hskip -0.2em} }
		\toprule
		\textbf{ID} & 
		\textbf{Categories} & 
		\textbf{Constraint(s)} \\ 
		\midrule
		$A$& 		$\userconstraints_I$&  
		$|g.role| \leq 3$ \\
		$M$ & 		$\userconstraints_I$& 
		$\texttt{sum}(g.duration) \geq 10^1$ \\
	   $N$ &  		$\userconstraints_I$& 
		$\texttt{avg}(g.duration) \leq 5*10^5$ \\
		$Gr$& 	$\userconstraints_G$& 
		$|G| \leq 3$ \\
		\midrule
		$C1$ & 	$\userconstraints_I, \userconstraints_G$ &
		$A \land N \land Gr$  \\
		$C2$ & 	$\userconstraints_I, \userconstraints_G$ & 
		 $A \land M \land N \land Gr$ \\
		\midrule
		$BL1$ & 	$\userconstraints_C$ & 
		 $|g| \leq 5$ \\
		$BL2$ & 	$\userconstraints_C$ & 
		$BL1$ $\land$ \texttt{cannotLink}($e_1.C,e_2.C$) \\
		$BL3$ & 	$\userconstraints_C$ &
		$|g.D| = 1$ \\
		$BL4$ & 	$\userconstraints_G$ & 
		$|G| = \nicefrac{|L|}{2}$\\
		\bottomrule
	\end{tabular}
\end{table}

%

\mypar{Configurations}
We test three configurations that differ in the instantiation of Step 1 of \approach (cf., \autoref{sec:approach:step1}): 
\begin{compactitem}
	\item \textbf{\configexh}, using exhaustive candidate computation;
	
	\item \textbf{\configdfg}, using the \dfg-based instantiation without beam search (i.e., unlimited beam width);
	
	\item \textbf{\configdfgbeam}, using the \dfg-based instantiation with a beam width that adapts to the number of event classes in the given log, i.e., $k$ = $5*|\eventclassesinlog|$.
\end{compactitem}

\noindent Note that we let candidate computation time out after 5 hours. \approach then continues with the candidates identified so far.

\mypar{Baselines}
We compare  \approach against three baselines. These represent alternative approaches to solve the log-abstraction problem and differ in the scope of constraints they can handle.\footnote{More details on the baselines and their implementation can be found in our repository linked in \autoref{sec:setup}}

		\mypartwo{Graph querying (\textbf{\blquery})} \approach's \dfg-based candidate computation traverses a DFG to find candidate groups that adhere to imposed constraints. 
		Recognizing the overlap of this with graph querying, \blquery replaces replaces Step~1 of \approach with an instantiation using graph querying.
		For this, the \dfg is stored in a graph database, which is queried for candidate groups using constraints formulated in a state-of-the-art graph querying language~\cite{cypher18}. 
		Given that a \dfg captures a log on the class-level, 
		\blquery can only support class-based constraints, though.
		Thus, we assess \blquery using a constraint on the maximum group size ($BL1$), an additional cannot-link constraint between event classes ($BL2$), and a constraint over a class-level attribute ($BL3$).
		By comparing against \blquery, we aim to show that \approach  yields more comprehensive sets of candidate groups than those obtained by adopting existing solutions.
	
		\mypartwo{Graph partitioning (\textbf{\blpart})} \approach's goal to find a disjoint set of cohesive groups for log abstraction is similar to the goal of graph partitioning, which aims to partition a graph such that edges between different groups have a low weight~\cite{luxburg2007spectral}. 		
		Therefore, we compare \approach against a baseline using such partitioning, \blpart. Given a  \dfg, \blpart aims to minimize the sum of directly-follows frequencies of cut edges, while cutting the graph into $n$ partitions. For this, \blpart applies \textit{spectral partitioning}~\cite{luxburg2007spectral}, where the weighted adjacency matrix is populated using normalized directly-follows frequencies. 
		Since graph partitioning simply splits a DFG into a certain number of groups, \blpart can only support strict grouping constraints, whereas instance-based, class-based, and flexible grouping constraints (e.g., constraint $Gr$), cannot be handled.
		Therefore, we compare \blpart against \approach using the constraint $BL4$, which aims to reduce number of event classes of a log by half.
		This comparison aims to show that \approach's three-step approach  leads to better log-abstraction results, while also supporting a considerably broader range of constraints.
	
		\mypartwo{Greedy approach (\textbf{\blgreedy})} 
		Finally, we compare \approach against a greedy abstraction strategy.
		\blgreedy starts by assigning all event classes from $\eventclassesinlog$ to a set of singleton groups, $G_0$. Then, in each iteration, \blgreedy merges those two groups from $G_i$ that lead to the lowest overall distance, i.e., $\dist(G_{i+1}, L)$, without resulting in any constraint violations. 
		\blgreedy stops if the overall distance cannot improve in an iteration.
	Unlike the other baselines, \blgreedy can handle instance-based constraints, since it works directly on the event log rather than the DFG, although, grouping constraints cannot be enforced in this iterative strategy. 
	Therefore, we compare \blgreedy against \approach using the instance-based constraint set  $A$, $M$, and $N$.
 	This comparison against \blgreedy aims to show the importance of striving for a global optimum in the log-abstraction problem.

\mypar{Measures}
To assess the results obtained by the various configurations and baselines, we consider the following measures:

\mypartwo{Solved abstraction problems (\textbf{\solved})}: 
	We report on the fraction of solved problems, to reflect the general feasibility of abstraction problems and the ability of a specific configuration to find such feasible solutions.
	
	\mypartwo{Size reduction (\textbf{\sizered})}: We measure the obtained size reduction by comparing the number of high-level activities in an obtained grouping  to the number of original event classes, i.e., $\nicefrac{|\finalgrouping|}{|C_L|}$. Given the strong link between model size and process understandability~\cite{reijers2010}, this measure provides a straightforward but clear quantification of the abstraction degree.
	
	\mypartwo{Complexity reduction (\textbf{\complred)}}: We also assess the abstraction degree through the reduction in \emph{control-flow complexity}, using an established complexity measure~\cite{reijers2010}. Since this measure requires a process model as input, we discover a model for both the original and the abstracted log using the state-of-the-art \emph{Split Miner}~\cite{augusto2019} and then compare their complexity.
		
	\mypartwo{Silhouette coefficient (\textbf{\silhouette})}: We quantify the  intra-group cohesion and inter-group separation of a grouping $\finalgrouping$ using the \emph{silhouette coefficient}~\cite{kaufman2009finding}, an established measure for cluster quality.
	To avoid bias,
	we compute this coefficient using a standard measure for the pair-wise distance between event classes~\cite{guenther07}, which considers their average positional distance.
	
 \mypartwo{Runtime (\textbf{\runtime})}: Finally, we measure the time in minutes required to obtain an abstraction result, from the moment a log $L$ is imported until the abstracted log $L'$ is returned.

\subsection{Evaluation Results}
\label{sec:realresults}
This section reports on the results for the different constraint sets, followed by a comparison of the different configurations. 

\mypar{Overall results}
\autoref{table:results:perconstraint} presents the results obtained using the \configexh configuration of \approach per constraint set.
For the anti-monotonic ($A$, $BL1$-$3$) and grouping constraint sets ($Gr$,$BL4$)  \approach finds a solution to all  of the problems. 
Infeasible problems primarily occur for the monotonic $M$ constraint set 
and the combination sets, $C1$ and $C2$,
 since these are more restrictive. 
 Interestingly, $C1$ has more than twice as many solved problems (54\%) than $C2$ (23\%), 
clearly showing the impact of $C2$'s additional monotonic constraint on feasibility.

The other measures in the table report on the results obtained for the solved problems. We observe that \approach achieves a considerable degree of abstraction, reflected in the  reductions in size and complexity. Groupings are reasonably cohesive and well separated from each other, indicated by silhouette coefficients $\geq$ 0.12. These results are stable for the less restrictive constraint sets, such as $A$, $N$, and $Gr$, as well as their combination $C1$. For instance, for $A$ a size reduction of 0.68, complexity reduction of 0.63, and silhouette coefficient of 0.15 is achieved. In line with expectations, for more restrictive constraint sets, e.g., \emph{C2}, the impact of abstraction is less significant (0.50, 0.40, and 0.09 resp.).
Finally, the impact of the constraint-checking modes on efficiency can also be observed.\footnote{
	For constraint sets with unsolved problems, runtimes must be compared carefully, as they strongly depend on the specific logs with feasible solutions.}  For instance, while the \emph{Gr} constraint set
requires 144m on average to be solved,  the anti-monotonic $BL2$ constraint cases are solved in 121m. 
In this mode candidates do not have to be expanded if they already violate the constraint, which leads to improved runtimes.

Overall, \approach is thus able to greatly reduce the size and complexity of event logs, while respecting various constraints. Although the  solution feasibility and the abstraction degree depends on the employed constraints, \approach consistently finds 
groups that have strong cohesion and good separation.

\begin{table}[!htb] 
	\small
	\centering
	\caption{Results for \configexh, averaged over solved problems.}
	\label{table:results:perconstraint}
	\vspace{-.3em}
	\begin{tabular}
		{c ccccc}
		\toprule
		\textbf{Const.}	&
		\multicolumn{1}{c}{\textbf{\solved}}&
		\multicolumn{1}{c}{\textbf{\sizered}}&
		\multicolumn{1}{c}{\textbf{\complred}}&
		\multicolumn{1}{c}{\textbf{\silhouette}}&
		\multicolumn{1}{c}{\textbf{\runtime}}\\
		\midrule
		\textbf{$A$} & 1.00 & 0.68 & 0.63 & 0.15 & 146\\
		\textbf{$M$} & 0.31 & 0.58 & 0.55 & 0.15 & 75\\
		\textbf{$N$} & 0.77 & 0.68 & 0.65 & 0.12 & 154\\
		\textbf{$Gr$} & 1.00 & 0.66 & 0.61 & 0.13 & 144\\
		\textbf{$C1$} & 0.54 & 0.68 & 0.59 & 0.12 & 134\\
		\textbf{$C2$} & 0.23 & 0.50 & 0.40 & 0.09 & 100\\	
		\textbf{$BL1$} & 1.00 & 0.67 &  0.61  & 0.12 & 122\\
		\textbf{$BL2$} & 1.00 & 0.66 &  0.61 & 0.12 & 121\\
		\textbf{$BL3$} & 1.00 & 0.38 & 0.29 & -0.02 & 38\\
		\textbf{$BL4$} & 1.00 & 0.51 & 0.46 & 0.05 & 147\\
		\bottomrule
	\end{tabular}
\end{table}

\mypar{Exhaustive versus efficient configurations}
\autoref{table:results:configs} depicts the evaluation results for the three \approach configurations, again providing the averages over the solved problems. 
Notably, the configurations were able to solve the same problems, except for a single problem in the non-monotonic $N$ constraint set, which the \configdfgbeam configuration failed to solve.

We observe that the \dfg-based configurations achieve substantial efficiency gains in comparison to the exhaustive one, where in particular \configdfgbeam needs only about 40\% of the time in comparison to \configexh (49m vs. 130m on average).

With respect to the abstraction degree, we observe that \configdfg maintains results  comparable to \configexh for size (0.62 vs. 0.63) and complexity reduction (0.56 vs. 0.57). It even obtains better results for the silhouette coefficient (0.16 vs. 0.11), which shows the ability of the \dfg-based approach to find candidate groups that are cohesive and well-separated. The results achieved by \configdfgbeam suggest a trade-off between optimal abstraction and efficiency, as the abstraction degree is about 7\% lower compared to the other configurations.

Finally, we observe that the DFG-based configurations are particularly useful for anti-monotonic and grouping constraints. In these cases, the results differ only marginally, even for \configdfgbeam, while achieving considerable efficiency gains.

\begin{table}[!htb] 
	\small
	\centering
	\caption{Results per configuration over solved problems.}
	\label{table:results:configs}
	\begin{tabular}
		{l ccccc}
		\toprule
	\textbf{Conf.}	&
		\multicolumn{1}{c}{\textbf{\solved}}&
		\multicolumn{1}{c}{\textbf{\sizered}}&
		\multicolumn{1}{c}{\textbf{\complred}}&
		\multicolumn{1}{c}{\textbf{\silhouette}}&
		\multicolumn{1}{c}{\textbf{\runtime}}\\
		\midrule
		\textbf{\configexh}& 0.78 & 0.63 & 0.57 & 0.11 & 130\\
		\textbf{\textsc{dfg}}$_\infty$& 0.78 & 0.62 & 0.56 & 0.16 & 108\\
		\textbf{\textsc{dfg}}$\mathit{_{k}}$& 0.77 & 0.56 & 0.50 & 0.08 & 49\\
		\bottomrule
	\end{tabular}
\end{table}

 \subsection{Baseline results}
 \label{sec:baselineresults}
 \autoref{table:results:bl} depicts the results obtained using the baseline approaches 
 against the most relevant configurations of \approach.

\mypar{Comparison to graph querying}
 	The results of \blquery indicate that the candidate groups obtained using graph queries are not as comprehensive as those found by \approach's \configdfg configuration. \blquery's solutions are therefore subpar with respect to size and complexity reduction.
	 Furthermore, the negative silhouette coefficients (-0.2 avg. vs. 0.17 for \configdfg) indicate that
	   the groupings found by \blquery are neither cohesive nor separated, which highlights the ability of
	   	 \dfg-based candidate computation to find better  sets of candidates for abstraction.
 
\mypar{Comparison to graph partitioning}
 	With respect to \blpart we find that partitioning the \dfg by minimizing edge cuts naturally reduces the size of the \dfg and, thus, achieves a certain degree of abstraction. However, the groupings created by \blpart are not as cohesive, indicated by the silhouette coefficient of 0.01 compared to \approach (0.05). Moreover, the complexity reduction achieved by \blpart (0.41) is lower than achieved by \approach (0.46), even though their groupings contain the  same number of activities. This highlights the benefits of the three-step approach \approach takes and the suitability of its distance measure to obtain meaningful groupings for log abstraction.

 \mypar{Comparison to greedy approach}
When considering the results of \blgreedy, the downsides of a greedy solution strategy quickly become apparent. \blgreedy finds solutions to fewer abstraction problems (64\%) than even the most efficient configuration, \configdfgbeam, whereas the solutions that are identified are  far subpar. For example, for the anti-monotonic $A$ constraint set, \blgreedy achieves an average size reduction of 0.47, whereas \configdfgbeam yields a size reduction of 0.64, which clearly shows that a greedy strategy often yields solutions that are far from optimal.
 
 \begin{table}[!htb] 
 	\small
 	\centering
 	\caption{Baseline comparison over the applicable constraint sets.  Results are averaged over solved problems.}
 	\label{table:results:bl}
 	\begin{tabular}
 		{ll cccc@{\hskip -0.1em}c}
 		\toprule
 		\textbf{Const.}&\textbf{Conf.}&
 		\multicolumn{1}{c}{\textbf{\solved}}&
 		\multicolumn{1}{c}{\textbf{\sizered}}&
 		\multicolumn{1}{c}{\textbf{\complred}}&
 		\multicolumn{1}{c}{\textbf{\silhouette}}&
 		\multicolumn{1}{c}{\textbf{\runtime}}\\
 		
 		\midrule
 		\multirow{2}{*}{\textbf{$BL$[$1$-$3$]}}
 		&\textbf{\textsc{dfg}$_\infty$}& 1.00& 0.63& 0.55 & \hphantom{-}0.17 & \hphantom{0001}77\\
 		&\textbf{\blquery}&0.96 & 0.55 & 0.43 & -0.20 & \hphantom{0001}24\\
 		\midrule
 		\multirow{2}{*}{\textbf{$BL4$}}
 		&\textbf{\configexh}&1.00 & 0.51 & 0.46 & \hphantom{-}0.05 &\hphantom{000}147 \\
 		&\textbf{\blpart}& 1.00& 0.51& 0.42 & \hphantom{-}0.01 & \hphantom{00047}1\\
 		\midrule
 		\multirow{2}{*}{\textbf{$A,M,N$}}
 		&\textbf{\textsc{dfg}$_k$}& 0.67& 0.59& 0.52 & \hphantom{-}0.08 & \hphantom{0001}58\\
 		&\textbf{\blgreedy}&0.64 & 0.45 & 0.37 & \hphantom{-}0.02 & \hphantom{0001}24\\
 		\bottomrule
 	\end{tabular}
 	 
 \end{table}

 \mypar{Discussion} Overall,  these results demonstrate that \approach outperforms all three baselines with respect to their applicable constraints, whereas it can, furthermore, handle a much broader range of process-oriented abstraction constraints.

\subsection{Case Study}
\label{sec:casestudy}
In this section we apply \approach in a case study to give an illustration of the value of constraint-driven log abstraction.

We use an event log~\cite{bpi17} capturing a loan application process at a large financial institution. 
Although the log only contains 24 event classes, its complexity is considerable, as evidenced by its 160 DFG edges. As shown in \autoref{fig:spaghetti} this issue even remains for a so-called 80/20 model, which omits the 20\% least frequent edges, since this visualization still provides few 
useful insights into the underlying process.

We recognize that the needed log abstraction can here be guided 
by considering the three IT systems from which events in the log originate: an \emph{application-handling} system (A), the \emph{offer} system (O), and a general \emph{workflow} system (W). Since these systems each relate to a distinct part of a process, we impose a constraint that avoids mixing up events from different systems into a single activity, i.e.,  $|g.origin| \leq 1$.

\begin{figure}[!htb]
	\centering
	\includegraphics[width=\linewidth]{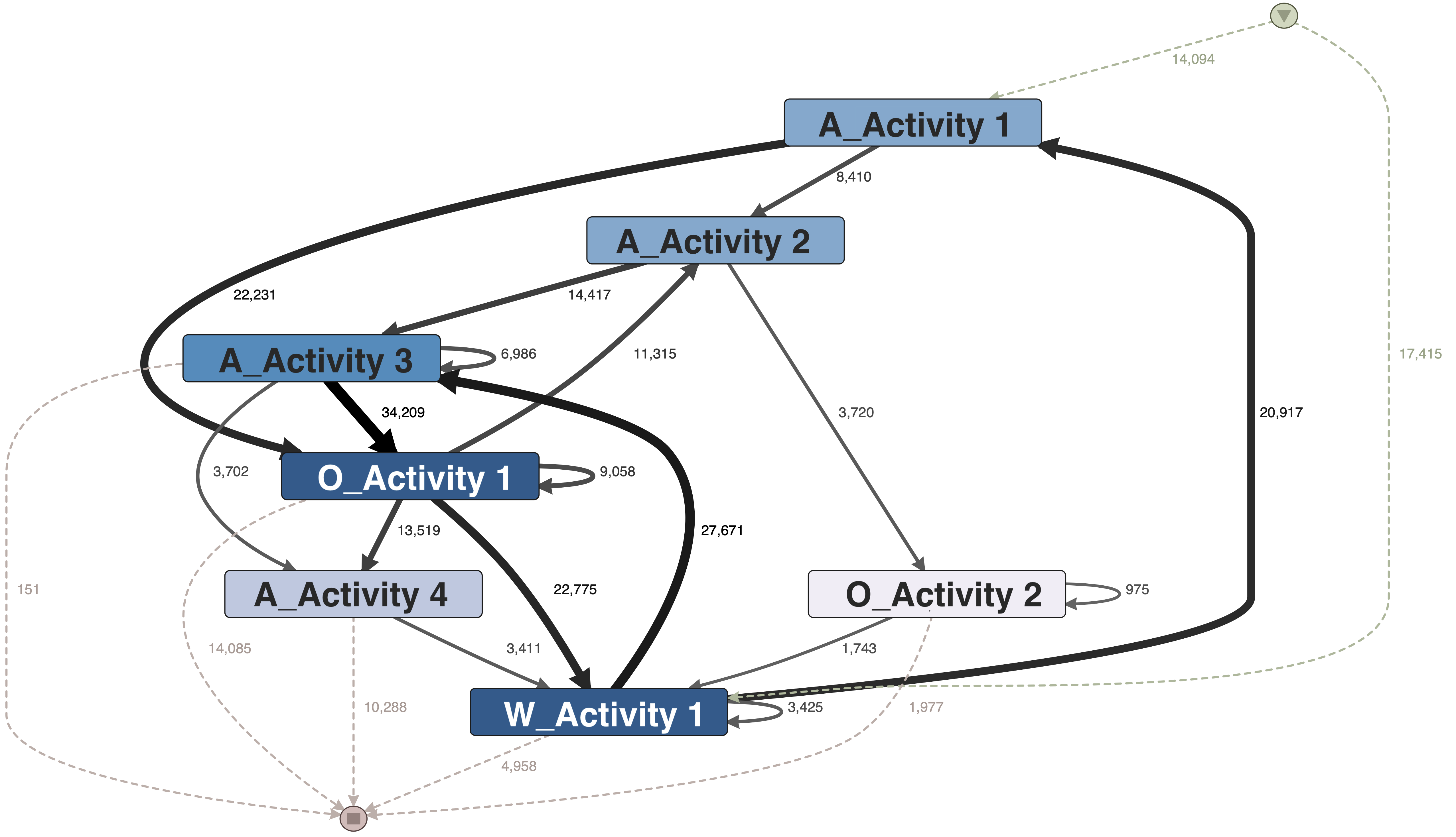}
	\caption{80/20 \dfg~of the abstracted loan application log.}
	\label{fig:loanabstracted}
\end{figure}

\autoref{fig:loanabstracted} depicts the DFG obtained by \approach in this manner, where the activity labels reflect their origin systems. 
 Having grouped the events into seven high-level activities, the DFG shows a considerable reduction in terms of  size and complexity. Due to this simplification, 
we observe clear inter-relations between the different sub-systems. For instance,  the process most often starts with the execution of steps in the application-handling system (\emph{A\_Activity 1} to \emph{3}), followed by a part in the offer system (\emph{O\_Activity 1}), and again concluded in the application-handling system (\emph{A\_Activity 4}). 
Next to this main sequence, the workflow-related steps (\emph{W\_Activity 1}) occur in parallel to the other activities, whereas the refusal of an offer (\mbox{\emph{O\_Activity 2}}) represents a clear alternative path.

It is important to stress that such insights are only possible due to the constraint-driven nature of our work. In fact, when applying \approach without imposing any constraints, the intertwined nature of the process even yielded high-level activities that contain events from all three sub-systems, thus obfuscating the key inter-relations in the process.



%% file: sections/related.tex

Our work primarily relates to the following streams:


\mypar{Log abstraction in process mining}
In the context of process mining, a broad range of techniques have been developed for log abstraction, also referred to as \emph{event abstraction}, of which Van Zelst et al.~\cite{VanZelst2020} and Diba et al.~\cite{Diba2020} recently provided comprehensive overviews.
Unsupervised techniques mostly employ clustering~\cite{folino2015mining,rehse2018clustering} or generic abstraction patterns~\cite{bose07,Wiegand2021} to group low-level events into high-level activities.
Other techniques are supervised, requiring users to provide information about the high-level activities to be discovered, captured ,e.g., in the form of a process model~\cite{baier2014bridging}, specific event annotations~\cite{Leemans20}, or domain hierarchies~\cite{klessascheck}.

Compared to \approach, existing unsupervised techniques do not guarantee any characteristics for the abstracted logs, which can result in a considerable loss of information (cf.,~\autoref{sec:casestudy})
, while the supervised techniques require a user to explicitly specify \emph{how} abstraction should be performed, whereas our work only needs a specification of \emph{what} properties they desire.

\mypar{Behavioral pattern mining from event logs}
Behavioral pattern mining also lifts low-level events to a higher degree of abstraction by identifying interesting patterns in event logs, including constructs such as exclusion, loops, and concurrency. 
Local Process Models (LPMs)~\cite{Tax2016} provide an established foundation for this. LPMs are mined according to pattern frequency, while extensions have been proposed to employ interest-aware utility functions~\cite{Tax2018}  and  incorporate user constraints~\cite{Tax2018a}.
Behavioral pattern mining has been also addressed through the discovery of maximal and compact patterns in logs~\cite{Acheli2019} and their context-aware extension~\cite{acheli21}.

While their purposes are similar, a key difference between behavioral pattern mining and log abstraction is that the former \emph{cherry-picks} interesting parts from an event log, whereas the latter strives for comprehensive abstraction over an entire log.

\mypar{Sequential pattern mining}
Approaches for sequential pattern mining~\cite{agrawal1995mining, pei2004mining} identify interesting patterns in sequential data. As for LPMs, \emph{interesting} is typically defined as \emph{frequent}~\cite{han2007frequent}, while techniques for \emph{high-utility sequential pattern mining} also support utility functions specific to the data attributes of events~\cite{truong2019survey, yin2013efficiently, yin2012uspan}.
Cohesion, comparable to distance measures used in log abstraction, has also been applied as a utility measure for pattern mining in single long sequences~\cite{cule2009,Cule2016}.
Furthermore, research in \emph{constrained sequential pattern mining} primarily focuses on exploiting constraint characteristics such as monotonicity to improve efficiency~\cite{Pei2007}, which we leverage during the first step of our approach as well. Frequently the focus is on specific constraints, e.g., time-based gap constraints~\cite{agoa2017}. 

In contrast to \approach, pattern mining techniques do not consider concurrency and exclusion. Moreover, like for behavioral pattern mining, the focus is on the identification of individual patterns, while our work strives for global abstraction.


%% file: sections/conclusion.tex
This paper proposed the \approach approach for constraint-driven abstraction of event logs. 
With \approach, users can define the desired characteristics and requirements of abstracted logs in terms of constraints, thus enabling the meaningful and purpose-driven abstraction of low-level event data. 
We provide two primary instantiations of our approach, allowing users to trade-off computational efficiency and result optimality.
Our evaluation experiments using real-life event logs reveal that our work considerably outperforms baseline techniques, whereas a case study demonstrates its usefulness in practical settings. 

There are several promising directions for future research. 
We aim to extend the scope of our work with instance-based constraints over the entire grouping (rather than per group).
Furthermore, we aim to develop an approach to suggest interesting constraints to users for a given log.
Finally, we plan to lift our work to online settings, so that identified groupings are dynamically adapted to new arrivals in a stream.
